\documentclass[twocolumn, times, tighten]{aastex62}

\usepackage{graphicx}	
\usepackage{amsmath}	
\usepackage{amssymb}	
\usepackage[utf8]{inputenc}  
\usepackage[T1]{fontenc}      

\usepackage{multirow}
\usepackage{longtable}
\usepackage{enumitem}

\usepackage[]{xcolor}

\definecolor{mygreen}{cmyk}{0.75002,0,1,0}

\newcommand{\fermi}{{\it Fermi}-LAT}

\newcommand{\tc}[1]{\multicolumn{1}{c}{#1}}

\received{}
\revised{}
\accepted{}
\submitjournal{ApJ}

\shorttitle{Radio activities of EESBHs}
\shortauthors{Yang et al.}

\begin{document}

\title{Radio activity of supermassive black holes with extremely high accretion rates}

\correspondingauthor{Xiaolong Yang}
\email{yxl.astro@gmail.com}

\author[0000-0002-4439-5580]{Xiaolong Yang}
\affiliation{Kavli Institute for Astronomy and Astrophysics, 
Peking University, 
Beijing 100871, China}
\affiliation{Shanghai Astronomical Observatory, Key Laboratory of Radio Astronomy, Chinese Academy of Sciences, 200030 Shanghai, P.R. China}

\author[0000-0002-9728-1552]{Su Yao}
\affiliation{Kavli Institute for Astronomy and Astrophysics, 
Peking University, 
Beijing 100871, China}
\affiliation{National Astronomical Observatories, 
Chinese Academy of Sciences, 
Beijing 100012, China}

\author{Jun Yang}
\affiliation{Department of Space, Earth and Enviroment, Chalmers University of Technology, Onsala Space Observatory, SE-439\,92 Onsala, Sweden}
\affiliation{Shanghai Astronomical Observatory, Key Laboratory of Radio Astronomy, Chinese Academy of Sciences, 200030 Shanghai, P.R. China}

\author{Luis C. Ho}
\affiliation{Kavli Institute for Astronomy and Astrophysics, Peking University, Beijing 100871, China}
\affiliation{Department of Astronomy, School of Physics, Peking University, Beijing 100871, China}

\author[0000-0003-4341-0029]{Tao An}
\affiliation{Shanghai Astronomical Observatory, Key Laboratory of Radio Astronomy, Chinese Academy of Sciences, 200030 Shanghai, P.R. China}

\author{Ran Wang}
\affiliation{Kavli Institute for Astronomy and Astrophysics, Peking University, Beijing 100871, China}
\affiliation{Department of Astronomy, School of Physics, Peking University, Beijing 100871, China}

\author{Willem A. Baan}
\affiliation{Xinjiang Astronomical Observatory, Key Laboratory of Radio Astronomy, Chinese Academy of Sciences, 150 Science 1-Street, 830011 Urumqi, P.R. China}
\affiliation{Netherlands Institute for Radio Astronomy ASTRON, NL-7991 PD Dwingeloo, the Netherlands}

\author{Minfeng Gu}
\affiliation{Shanghai Astronomical Observatory, Key Laboratory for Research in Galaxies and Cosmology, Chinese Academy of Sciences, 200030 Shanghai, P.R. China}

\author{Xiang Liu}
\affiliation{Xinjiang Astronomical Observatory, Key Laboratory of Radio Astronomy, Chinese Academy of Sciences, 150 Science 1-Street, 830011 Urumqi, P.R. China}

\author{Xiaofeng Yang}
\affiliation{Xinjiang Astronomical Observatory, Key Laboratory of Radio Astronomy, Chinese Academy of Sciences, 150 Science 1-Street, 830011 Urumqi, P.R. China}

\author{Ravi Joshi}
\affiliation{Kavli Institute for Astronomy and Astrophysics, Peking University, Beijing 100871, China}



\begin{abstract}


Radio emission from the high- and super-Eddington accreting active galactic nuclei (AGNs) has various origins: a persistent jet, the magnetized corona and the wind-like outflows. It is now still unclear which is the leading mechanism responsible for the observed radio emission and how the radio emission is related to other characteristic parameters such as the Eddington ratio and black hole mass. In this paper, we present the 5-GHz Very Large Array (VLA) observational results of a sample of 25 extremely high Eddington accreting supermassive black holes (EESBHs, the Eddington ratio $\lambda_\mathrm{Edd}$ close to or above one) from narrow-line Seyfert 1 galaxies, among which 22 sources are detected. Most of EESBHs show a compact radio structure from a few hundred parsecs to one-kiloparsec scale. We estimated the lowest star formation rate surface density required for producing the observed radio emission, and found it is higher than the largest value previously detected in circumnuclear starburst galaxies, implying that the radio emission is from the AGN activity. Along with a comparison sample, we find an overall inverse $\mathcal{R}$ - $\lambda_\mathrm{Edd}$ correlation ranging from sub- to super-Eddington ratios. 
The high-Eddington and mildly super-Eddington AGNs ($-0.5<$log\,$\lambda_\mathrm{Edd}<0.6$) have a radio to X-ray luminosity ratio $L_\mathrm{R}/L_\mathrm{X}\sim10^{-5} $--$ 10^{-4}$ and a steep radio spectrum, supporting that the radio emission is from transient ejecta (outflows) of corona, however, the jet contribution cannot be fully ruled out. Our highly super-Eddington sources (log\,$\lambda_\mathrm{Edd}\gtrsim0.6$) have a flatter radio spectrum, along with its low radio luminosity: $L_\mathrm{R}/L_\mathrm{X}\sim10^{-5}$, their radio emission is likely dominated by a magnetized corona, and a radiation pressure caused jet is also proposed in this paper.

\end{abstract}



\section{Introduction} \label{sec:intro}

With increasing observational evidence during the past two decades, it is now widely accepted that supermassive black holes (SMBHs) with $M_{\rm BH}\sim10^{5-10}\,M_{\odot}$ reside at the centers of most galaxies \citep[see review in][]{2013ARA&A..51..511K}. The growth of SMBHs is believed to primarily occur through gas accretion, which gives rise to a luminous active galactic nucleus (AGN) with a large energy release in the form of radiation. The total energy radiated by this process cannot exceed the Eddington luminosity for the spherically symmetric accretion, which was thought to be the main mechanism for regulating the growth of SMBHs. However, observations have shown that super-Eddington accretion can occur in a variety of celestial systems, such as Galactic X-ray binaries (XRBs), \textit{e.g.}, SS\,433 \citep[][]{2002ApJ...566.1069G, 2006MNRAS.370..399B, 2015NatPh..11..551F, 2018arXiv181010518M} and GRS\,1915+105 \citep[][]{1994Natur.371...46M, 2001Natur.414..522G}; tidal disruption events \citep[TDEs, e.g.][]{2011Sci...333..203B, 2011Natur.476..421B, 2011Sci...333..199L, 2011Natur.476..425Z, 2017ApJ...838..149A, 2018MNRAS.478.3016W, 2018ApJ...859L..20D} and narrow-line Seyfert\,1 galaxies \citep[NLS1s, e.g.][]{2016A&A...590A..77L}. Moreover, the violation of the Eddington limit becomes increasingly important for the various related topics of black hole growth, galaxy evolution and AGN feedback, \textit{e.g.}, in ultra-luminous X-ray sources \citep[ULXs, see][and references therein]{2017ARA&A..55..303K} with stellar-mass black hole/neutron stars, and primordial massive black holes during their early rapid growth  \citep[e.g.][]{2005ApJ...633..624V, 2015ApJ...804..148V, 2018MNRAS.476..673T}.

The accretion states and their transitions in XRBs are now well studied by using the hardness-intensity diagrams (HIDs), which are also coupled with radio emission (and hence the jets) and the Eddington ratios. It has been proposed frequently that AGNs experience accretion state transitions \citep[e.g.][]{2006MNRAS.372.1366K} in a similar way with XRBs in the framework of AGN-XRB unification \citep{2004A&A...414..895F} but with substantially longer timescales \citep[$>10^5$\,yr, e.g.][]{2015MNRAS.451.2517S}. It is thus possible to connect these compact accreting systems by using scaling relations in parameter space with the most fruitful scheme of the triple correlation among X-ray luminosity, radio luminosity, and black hole mass \citep[i.e., the so-called fundamental plane relation,][]{2003MNRAS.345.1057M}. However, the fundamental plane relation is found to be successful only in unifying low luminosity AGNs and XRBs in the hard state. 

The continuous jets are ubiquitous at low accretion rates (the low/hard state) in XRBs, but intermittent or entirely absent at high accretion rates (the high/soft state and the intermittent/very high state) \citep[e.g.][]{1996ApJ...459..185M, 1999ApJ...519L.165F, 2004MNRAS.355.1105F}. Resembling the inverse correlation between the radio luminosity of jets and X-ray luminosity in XRBs,  \citet{2002ApJ...564..120H} found a similar inverse correlation between radio loudness ($\mathcal{R}\equiv L_{\nu_5}/L_{\nu_B}$) and the Eddington ratio ($\lambda_\mathrm{Edd} \equiv L_\mathrm{bol}/L_\mathrm{Edd}$) in AGNs. In this scheme, radio-loud AGNs with powerful relativistic jets often have low Eddington ratios, and vice versa. In contrast to the fundamental plane relation only valid for certain conditions, the inverse correlation between $\mathcal{R}$ and $\lambda_\mathrm{Edd}$ is ubiquitous in both AGNs and XRBs \citep[e.g.][]{2011MNRAS.417..184B} though with a large scatter. A global analogy between stellar-mass black holes and supermassive black holes has been established in the $\mathcal{R}$ and $\lambda_\mathrm{Edd}$ correlation: low-luminosity AGNs are similar to XRBs in the low/hard state, and the high- or super-Eddington accreting AGNs (e.g., narrow-line Seyfert 1s) being an analogy of XRBs in the high/soft and the very high state. It should be noted here that only a few XRBs experience transitions from classical spectral states to the super-Eddington regime \citep[e.g.][]{2009Natur.458..481N}. The super-Eddington accretion state is poorly understood primarily because of the extremely short timescales of the spectral state transition in XRBs. Therefore, the study of super-Eddington accretion and the associated jet formation and/or quenching in XRBs does not warrant a scaling-up to AGNs.

The dynamic timescale in different accretion states is proportional to the mass of the central black hole, and it is difficult to observe a complete burst cycle in individual AGNs resembling XRBs. Still, progress has been made by \citet{2007ApJ...658..815S}, providing further confirmation of $\mathcal{R}-\lambda_\mathrm{Edd}$ inverse correlation with the Eddington ratio range from sub- to super-Eddington ratios in both radio-loud and radio-quiet AGNs. In this paper, the short-lived super-Eddington accreting AGNs will be studied to augment the correlation in the super-Eddington regime with robust measurements of the radio loudness and the Eddington ratio.

\citet{2006ApJ...636...56G} have found that AGNs with super-Eddington accretion rates are predominately radio-quiet. The radio emission from radio-quiet AGNs can have various origins, including persistent jets, a magnetized corona/jet base and a wind-like outflow, in this paper, the "jets" specially refer to a collimated outflow. Furthermore, star-forming activities in the host galaxy may also contribute to the thermal and non-thermal radio emission, which typically shows host-like extension, with a diffused and clumpy structure, and has a low surface brightness. Nuclear starbursts have been observed in some galaxies \citep[e.g.][]{2006AJ....132..321D, 2018MNRAS.477.1086H}, and a significant fraction of the narrow-line Seyfert 1 galaxies (NLS1s) tend to have circumnuclear star-forming rings \citep{2006AJ....132..321D}, making them more difficult to distinguish from the AGN. These different mechanisms are fundamentally crucial in super-Eddington accreting AGNs, and in explaining the $\mathcal{R} - \lambda_\mathrm{Edd}$ correlation. 

As mentioned above, the origin of the radio emission from the super-Eddington accreting AGN itself is complicated. The slim-disk model was firstly proposed by \citet{1988ApJ...332..646A} to describe a super-Eddington accretion flow. Recently, progress has been made in simulating the super-Eddington accretion disk \citep[e.g.][]{2011MNRAS.413.1623D, 2017MNRAS.464.1102B, 2017arXiv170902845J} and several models among them support the launching of a jet, which is driven by either the radiation-pressure \citep{2009PASJ...61..783T, 2015MNRAS.453.3213S} or the magnetic field retrieved from the spin of the central black hole, i.e., a Blandford-Znajek jet \citep[BZ,][]{1977MNRAS.179..433B} \citep[e.g.][]{2003PASJ...55L..69N, 2014MNRAS.445.3919K, 2015MNRAS.454L...6M}. From observations, several super-Eddington accreting systems are found to accelerate a jet, the trigger has been attributed to either a radiation-pressure \citep[e.g.][]{2011ApJ...736....2O, 2015MNRAS.453.3213S} or the BZ magnetic field \citep[e.g.][]{2012ApJ...748...36B}. Furthermore, a super-Eddington accretion disk may also drive a strong wind and accelerate relativistic electrons to produce the observed synchrotron radio emission. For example, in SS\,433, such wind-like radio-emitting outflows co-exist with a relativistic jet, and they have a comparable radio luminosity \citep[e.g.][]{2001ApJ...562L..79B}. In AGN accretion models, a hot $T_b\sim10^9$\,K corona in the vicinity of the black hole accretion disk may explain the observed X-ray emission from AGNs \citep{1991ApJ...380L..51H}. It has been suggested that a magnetically heated corona may also produce the radio emission \citep{2008MNRAS.390..847L, 2016MNRAS.459.2082R}. This hypothesis is later supported by the discovery of the similar radio to X-ray luminosity ratio $L_\mathrm{R}/L_\mathrm{X}\sim10^{-5}$ between radio-quiet quasars and coronally active stars \citep{2008MNRAS.390..847L}, and by high radio frequency ($\gtrsim100$\,GHz) observations of two nearby Seyfert galaxies \citep{2015MNRAS.451..517B, 2018MNRAS.478..399B, 2018ApJ...869..114I}.

Based on the topics mentioned above, the NLS1 galaxies may be introduced as a long-lived high Eddington ratio laboratory with an SMBH engine. NLS1s consist of a distinct class of AGNs identified by the width of their H$_\beta$ emission lines \citep[$<2000$\,km/s,][]{1985ApJ...297..166O, BG92}. They are located at an extreme end of the AGN parameter space that is believed to be governed mainly by the Eddington ratio \citep[e.g.][]{BG92}. Indeed, there is growing evidence showing that NLS1s have higher Eddington ratios and lower black hole masses than normal Seyfert 1 galaxies and quasars \citep[e.g.][]{BG92, 2015ApJ...806...22D}, which implies that they are systems with rapidly accreting SMBHs. NLS1s are often radio-quiet in contrast to the broad-line AGNs \citep[e.g,][]{1995AJ....109...81U}. Interestingly, NLS1s are not completely radio quiescent, and the origin of their weak radio emission holds the long controversy. Several papers have presented evidence for the presence of radio jets in a handful of radio-quiet NLS1s \citep[e.g.,][]{2004A&A...425...99L, 2009ApJ...706L.260G, 2013ApJ...765...69D, 2015ApJ...798L..30D}, which is far from a consensus. However, as a sub-class, the super-Eddington accreting NLS1s have not been comprehensively studied before. To study the super-Eddington accretion and also to investigate the origin of the radio emission in extremly high Eddington ratio accreting AGNs, we present high-resolution (sub-kpc-scale) VLA results of a sample of NLS1s that are accreting at Eddington ratios close to or exceeding one. 

This paper is organized as follows: In Section~\ref{sec:sample} we describe our sample; Section~\ref{sec:data} presents the procedure of data reduction; Sections~\ref{sec:results1} and \ref{sec:results2} provide results from data reduction and analysis, followed with the discussion in Section~\ref{sec:dis}. 
Throughout this paper we assume the same $\Lambda$-CDM cosmology with \citet{2007ApJ...658..815S} and \citet{2011MNRAS.417..184B}, which is $H_0=100\,h=70\,\mathrm{km}\,\mathrm{s^{-1}}\,\mathrm{Mpc^{-1}}, \Omega_M=0.3$, and $\Omega_{\Lambda}=0.7$.

\begin{table*}
\renewcommand\arraystretch{0.8}                                                                                                                                                                              
\centering                                                                                                                                                                                 
\caption{The information of our EESBHs} \label{tab:sample}                                                                                                                                                                                                                                                                                                                                                  
\begin{tabular}{cccccccc}                                                                                                                                                         
\toprule                                                                                                                                                                                   
Short Name   &    Alias                           &        z            &$   M_\mathrm{BH}         $&  log\,$f_\mathrm{(2-10\mathrm{keV})} $& log\,$ L_\mathrm{B}  $& log\,$L_\mathrm{bol}    $&$     \lambda_\mathrm{Edd}  $     \\
             &                                   &                     &($\times10^7$\,M$_\odot$)  &  (erg\,cm$^{-2}$\,s$^{-1}$)              &  (erg\,s$^{-1}$)   &    (erg\,s$^{-1}$)        &                                  \\
(1)          &    (2)                            &       (3)           &        (4)                &           (5)                         &        (6)         &            (7)            &             (8)                       \\
\hline                                                                                                                                           
J0230-0859   &    Mrk\,1044                      &$     0.016          $&$     0.18               $&$        -11.16                       $&$     42.91        $&$     45.08 - 46.04{}^a   $&$       5.14 -  46.81    $             \\                                                                                                                                                                                                                                                                                                                                                                                                                                                                                   
J2242+2943   &    Akn\,564                       &$     0.025          $&$     0.21               $&$        -10.62                       $&$     43.25        $&$     44.50{}^b           $&$       1.15             $             \\                                                                                                                                                 
J1153+4612   &    Mrk\,42                        &$     0.025          $&$     0.09               $&$        -11.61                       $&$     42.48        $&$    >43.48               $&$      >0.26             $             \\                                                                                                                                                  
J0006+2012   &    Mrk\,335                       &$     0.026          $&$     1.27               $&$        -10.73                       $&$     43.55        $&$     44.90 - 45.71{}^a   $&$       0.48 -   3.11    $             \\                                                                                                                                                   
J1559+3501   &    Mrk\,493                       &$     0.031          $&$     0.10               $&$        -11.42                       $&$     42.92        $&$     44.86{}^c           $&$       5.58             $             \\                                                                                                                                                  
J1703+4540   &    B3\,1702+457                   &$     0.060          $&$     0.38               $&$        -11.23                       $&$     43.47        $&$    >44.47               $&$      >0.60             $             \\                                                                                                                                                  
J1536+5433   &    Mrk\,486                       &$     0.039          $&$     0.90               $&$        -11.36                       $&$     43.32        $&$     44.42 - 45.23{}^a   $&$       0.22 -   1.45    $             \\                                                                                                                                                   
J1034+3938   &    KUG\,1031+398                  &$     0.042          $&$     0.25               $&$        -12.06                       $&$     43.24        $&$     45.01{}^c           $&$       3.15             $             \\                                                                                                                                                   
J1242+3317   &    IRASF\,12397+3333              &$     0.043          $&$     0.59               $&$        -11.24                       $&$     43.05        $&$     46.04 - 46.92{}^a   $&$      14.32 - 108.64    $             \\                                                                                                                                                   
J1025+5140   &    Mrk\,142                       &$     0.045          $&$     0.93               $&$        -11.13                       $&$     43.40        $&$     45.74 - 46.60{}^a   $&$       4.54 -  32.96    $             \\                                                                                                                                                   
J1431+2817   &    Mrk\,684                       &$     0.046          $&$     0.69               $&$        -11.58                       $&$     43.67        $&$    >44.67               $&$      >0.53             $             \\                                                                                                                                                   
J1246+0222   &    PG\,1244+026                   &$     0.048          $&$     0.19               $&$        -11.64                       $&$     43.39        $&$     45.48{}^c           $&$      12.24             $             \\                                                                                                                                                   
J0444+1221   &    IRAS\,04416+1215               &$     0.089          $&$     1.96               $&$        -11.80                       $&$     44.07        $&$     46.60 - 47.55{}^a   $&$      15.63 - 139.31    $             \\                                                                                                                                                  
J1014-0418   &    PG\,1011-040                   &$     0.058          $&$     2.12               $&$        -13.06                       $&$     44.15        $&$    >45.15               $&$      >0.51             $             \\                                                                                                                                                   
J0053+1241   &    IZw1                           &$     0.059          $&$     2.38               $&$        -11.07                       $&$     44.46        $&$     45.50 - 45.68{}^a   $&$       1.02 -   1.54    $             \\                                                                                                                                                   
J1451+2709   &    PG\,1448+273                   &$     0.065          $&$     1.03               $&$        -11.69                       $&$     44.09        $&$     45.83{}^c           $&$       5.05             $             \\                                                                                                                                                   
J0041+4021   &    Mrk\,957                       &$     0.071          $&$     0.14               $&$        -12.30                       $&$     43.27        $&$    >44.27               $&$      >1.02             $             \\                                                                                                                                                  
J0107+1408   &    SDSS\,J010712.04+140845.0      &$     0.076          $&$     0.09               $&$        -12.73                       $&$     42.75        $&$    >43.75               $&$      >0.49             $             \\                                                                                                                                                   
J1442+3526   &    Mrk\,478                       &$     0.077          $&$     2.32               $&$        -11.57                       $&$     44.40        $&$     45.64{}^d           $&$       1.44             $             \\                                                                                                                                                   
J1140+0307   &    SDSS\,J114008.71+030711.4      &$     0.081          $&$     0.09               $&$        -12.59                       $&$     42.94        $&$     45.07{}^c           $&$      10.04             $             \\                                                                                                                                                   
J1355+5612   &    RX\,J1355.2+5612               &$     0.122          $&$     0.93               $&$        -12.08                       $&$     43.94        $&$    >44.94               $&$      >0.72             $             \\                                                                                                                                                   
J1118+4025   &    PG\,1115+407                   &$     0.155          $&$     2.60               $&$        -11.81                       $&$     43.91        $&$     45.88{}^c           $&$       2.24             $             \\                                                                                                                                                   
J0207+0242   &    Nab\,0205+024                  &$     0.156          $&$     4.54               $&$        -11.30                       $&$     44.72        $&$    >45.72               $&$      >0.90             $             \\                                                                                                                                                   
J1405+2555   &    PG\,1402+261                   &$     0.164          $&$     3.93               $&$        -11.64                       $&$     44.08        $&$     46.02{}^c           $&$       2.05             $             \\                                                                                                                                                   
J2236+1343   &    PG\,2233+134                   &$     0.326          $&$     8.21               $&$        -12.29                       $&$     44.81        $&$     47.13{}^d           $&$      12.64             $             \\

\hline                                                                                                                                                                                                         
\end{tabular}

\begin{flushleft}                                                                                                                                                                                             
\textbf{Notes}: Column 1: short name; Column 2: alias; Column 3: redshift; 
Column 4: black hole mass estimated from broad emission line; Column 5: 2-10\,keV X-ray flux density;                                                                                            
Column 6: $k$-corrected $B$-band optical luminosity of AGNs (erg\,s$^{-1}$), calculated from 5100\,\AA~flux density $f_{5100}$,                                                                        
fraction of AGN contribution at 5100\,\AA~and an assumed optical spectral index $\alpha_{opt}=-0.5$; Column 7: bolometric luminosity (erg\,s$^{-1}$)                                                                              
and references; Column 8: Eddington ratio. 
                                                                                                                                                                       
\textbf{References}: (a) \citet{2016MNRAS.458.1839C}; (b) \citet{2013MNRAS.435.1840R};                                                                                                                        
(c) \citet{2012MNRAS.420.1825J}; (d) \citet{2012MNRAS.426.2677R};                                                                                                                                             
                                                                                                                                                                                                                                                                                                                                                                                                                                                                                                                                                                                                                                                                                                                                                                                                                                                                    
\end{flushleft}

\end{table*}                                                                                                                                                                                                  

\begin{table*}                                                                                    
\renewcommand\arraystretch{0.8}                                                                                                                                                                              
\centering                                                                                                                                                                                 
\caption{Summary of VLA A-array observations} \label{tab:vla}                                                                                                                                                                                                                                                                                                                                                 
\begin{tabular}{ccccrccrc}                                                                                                                                                         
\toprule                                                                                            
                                                                                                                                                                                   
Short Name   & Receiver &   Obs.\,ID   &      Date      &  Time     &  $\theta_{maj}$  &   $\theta_{min}$  &    P.A.       &    Refs.   \\
             &      &              &                &  (s)     &    (arcsec)             &         (arcsec)         &  (degree)     &            \\
(1)          & (2)  &    (3)       &      (4)       &  (5)      &      (6)                &            (7)           &    (8)        &    (9)     \\
                                                                                                    
\hline

J0230-0859            &  C-band   &    AU0028    &    1987-07-29  &  1310     &  0.70     &     0.39  &$   -32.0  $&   a     \\                  
                      &  L-band   &    AU0028    &    1987-07-29  &  690      &  2.30     &     1.27  &$   -32.0  $&   a     \\                  
                      &  X-band   &    AS0727    &    2002-02-16  &  1170     &  0.45     &     0.23  &$   -38.2  $&         \\                  
J2242+2943            &  C-band   &    15A-283   &    2015-07-02  &  598      &  0.50     &     0.43  &$    65.4  $&   b     \\                  
                      &  L-band   &    AL0502    &    1999-08-26  &  320      &  1.49     &     1.38  &$   -25.5  $&         \\                  
                      &  X-band   &    AB0973    &    2000-11-14  &  1310     &  0.23     &     0.21  &$    -4.0  $&   d     \\                  
J1153+4612            &  C-band   &    AU0028    &    1987-07-31  &  1790     &  0.50     &     0.40  &$    58.8  $&   a     \\                  
J0006+2012            &  C-band   &    15A-283   &    2015-07-05  &  538      &  0.44     &     0.38  &$    55.3  $&   b     \\                  
                      &  L-band   &    AM0384    &    1992-12-31  &  920      &  1.68     &     1.49  &$   -35.5  $&         \\                  
                      &  X-band   &    AM0384    &    1992-12-31  &  1050     &  0.25     &     0.23  &$   -40.0  $&         \\                  
J1559+3501            &  C-band   &    15A-283   &    2015-07-23  &  523      &  1.30     &     0.34  &$    59.9  $&   b     \\                  
                      &  L-band   &    AG0777    &    2008-12-05  &  1163     &  1.43     &     1.42  &$    67.6  $&         \\                  
                      &  X-band   &    12B-064   &    2012-10-25  &  3590     &  0.24     &     0.19  &$   -63.1  $&   f     \\                  
J1703+4540            &  C-band   &    15A-283   &    2015-07-22  &  538      &  0.73     &     0.39  &$    77.9  $&   b     \\                  
                      &  L-band   &    AM0492    &    1995-07-12  &  330      &  1.60     &     1.38  &$   -70.6  $&         \\                  
                      &  X-band   &    AM0492    &    1995-07-12  &  320      &  0.27     &     0.23  &$   -71.7  $&         \\                  
J1536+5433            &  C-band   &    15A-283   &    2015-08-29  &  658      &  0.66     &     0.37  &$    82.9  $&   b     \\                  
J1034+3938            &  C-band   &    15A-283   &    2015-08-29  &  598      &  0.47     &     0.40  &$    83.8  $&   b     \\                  
J1242+3317            &  C-band   &    15A-283   &    2015-09-06  &  553      &  0.45     &     0.36  &$    83.0  $&   b     \\                  
J1025+5140            &  C-band   &    AK0096    &    1983-11-20  &  1350     &  0.46     &     0.41  &$    50.7  $&   c     \\                  
J1431+2817            &  C-band   &    AU0028    &    1987-07-31  &  1800     &  0.49     &     0.42  &$   -37.8  $&   a     \\                  
J1246+0222            &  C-band   &    15A-283   &    2015-09-06  &  553      &  0.52     &     0.41  &$   -42.3  $&   b     \\                  
J0444+1221            &  L-band   &    AM0492    &    1995-07-10  &  340      &  1.50     &     1.28  &$     2.5  $&         \\                  
                      &  X-band   &    AM0492    &    1995-07-10  &  330      &  0.25     &     0.21  &$     7.3  $&         \\                  
J1014-0418            &  C-band   &    AK0096    &    1983-11-21  &  1470     &  0.66     &     0.38  &$   -37.3  $&   c     \\                  
J0053+1241            &  C-band   &    AC0624    &    2002-05-02  &  590      &  0.45     &     0.39  &$     2.7  $&         \\                  
                      &  L-band   &    AK0406    &    1995-07-20  &  370      &  1.55     &     1.32  &$    -3.2  $&         \\                  
                      &  X-band   &    AB0670    &    1992-12-04  &  7160     &  0.26     &     0.24  &$    24.4  $&         \\                  
J1451+2709            &  C-band   &    AK0096    &    1983-11-20  &  1430     &  0.45     &     0.41  &$   -36.4  $&   c     \\                  
J0041+4021            &  C-band   &    AH0221    &    1986-05-18  &  2070     &  0.48     &     0.38  &$   -10.9  $&         \\                  
                      &  L-band   &    AL0502    &    1999-08-26  &  330      &  1.44     &     1.40  &$    66.0  $&         \\                  
J0107+1408            &  C-band   &    AG0670    &    2004-10-10  &  1713     &  0.43     &     0.38  &$     8.3  $&   g     \\                  
J1442+3526            &  C-band   &    AK0096    &    1983-11-20  &  1420     &  0.43     &     0.42  &$   -69.5  $&   c     \\                  
                      &  L-band   &    AG0574    &    1999-09-25  &  315      &  1.24     &     1.16  &$   -59.5  $&   e     \\                  
                      &  X-band   &    AG0574    &    1999-09-25  &  430      &  0.24     &     0.23  &$    46.2  $&   e     \\                  
J1140+0307            &  C-band   &    AG0670    &    2004-10-10  &  2819     &  0.53     &     0.39  &$   -34.2  $&   g     \\                  
J1355+5612            &  C-band   &    15A-283   &    2015-08-30  &  553      &  1.14     &     0.38  &$    60.5  $&   b     \\                  
J1118+4025            &  C-band   &    AK0096    &    1983-11-20  &  1430     &  0.43     &     0.41  &$    71.6  $&   c     \\                  
J0207+0242            &  C-band   &    AH0333    &    1989-01-09  &  840      &  0.63     &     0.40  &$    41.0  $&         \\                  
                      &  L-band   &    WARD      &    1982-03-04  &  690      &  1.75     &     1.38  &$    32.5  $&         \\                  
J1405+2555            &  C-band   &    AK0096    &    1983-11-14  &  1470     &  0.42     &     0.39  &$   -15.9  $&   c     \\                   
                      &  L-band   &    AD0302    &    1992-12-11  &  730      &  1.46     &     1.34  &$   -22.1  $&         \\                  
J2236+1343            &  C-band   &    AK0096    &    1983-11-23  &  1530     &  0.51     &     0.40  &$   -44.2  $&   c     \\                  
                      &  L-band   &    AH0291    &    1987-10-11  &  300      &  1.49     &     1.28  &$   -15.2  $&         \\                  
                      &  X-band   &    AG0499    &    1996-10-29  &  3720     &  0.27     &     0.25  &$     4.6  $&         \\

\hline                                                                                                                                                                                  
\end{tabular}

\begin{flushleft}
\textbf{Notes}: Column 1: short name; Column 2: observational bands; Column 3: project ID; Column 4: observing date; 
Column 5: observing time of target source; Column 6 - 8: beam major axis, minor axis and position angle; Column 9: references 
for data sets publication.
                                                                                                                                                                     
\textbf{References}: (a) \citet{1995AJ....109...81U}; (b) \citet{2018A&A...614A..87B}; 
(c) \citet{1989AJ.....98.1195K}; (d) \citet{2001ApJS..132..199S}; (e) \citet{2000ApJ...537..152K}; 
(f) \citet{2014ApJ...788L..22G}; (g) \citet{2006ApJ...636...56G}.                                                                                                                                                                                                                                                                                                                                                                                                                                                                                                                                                                                                                                                                                                                                                                                                                                                             
\end{flushleft}

\end{table*}                                                                                                                                                                          

\section{The Sample} \label{sec:sample}

Our parent sample is comprised of 60 super-Eddington accreting supermassive black hole candidates with Eddington ratios close to or above one, which was compiled by \citet[][]{2013PhRvL.110h1301W}. 
Given a strong relation between the Eddington ratio $\lambda_\mathrm{Edd}$ and the X-ray photon index $\Gamma_{2-10\rm\,keV}$ \citep[e.g.][]{1999ApJ...526L...5L, 2013MNRAS.433.2485B}, AGNs with higher Eddington ratios are expected to have a steeper hard X-ray photon index $\Gamma_{2-10\rm\,keV}$ according to this well-known relation. The super-Eddington AGN candidates have actually been selected indirectly by having $\Gamma_{2-10\rm\,keV}>2$. This selection approach requires an accurate determination of their bolometric luminosities $L_{\rm bol}$ and the knowledge of the broadband spectral energy distribution \citep[SED, e.g.][]{2017MNRAS.471..706J}. Unfortunately, it is difficult to measure $L_{\rm bol}$ from broadband SED since much of the radiation from the accretion disk is in the form of extreme UV, which is beyond the observation windows. 
Still, accurate bolometric luminosity measurements (by using SED fitting) have been obtained from the literature for most of our super-Eddington AGN candidates (see column 7 of Table \ref{tab:sample} and the corresponding references) with a robust mass determination. 

To study the radio emission from super-Eddington AGNs, we have searched our parent sample for available high-resolution radio observations in the Very Large Array (VLA) data archive. We primarily collected VLA observations at C- and X-band with the A-array configuration having a resolution corresponding to a projected linear size $\lesssim2\rm\,kpc$ at the rest frame of each object. Furthermore, we also collected L-band observations with the VLA A-array when there has no high-resolution observational data at C and X bands. Twenty-six sources meet the requirements, which also includes an intrinsic radio-loud NLS1, 1H\,0323+342 \citep[][]{2007zhou} that has been detected in $\gamma$-rays by \fermi\ \citep[][]{2009ApJ...707L.142A}.  Observations show that the radio emission in this source is strongly Doppler boosted \citep[e.g.][]{2018arXiv180508299H}. This source was removed from our final sample of 25 objects. The basic information of the objects used for this research is listed in Table \ref{tab:sample}. The black hole mass and 2-10\,keV flux density are obtained from \citet[][]{2013PhRvL.110h1301W}, where the black hole mass was estimated by using the broad emission line width and the 5100\,\AA\,luminosity. We should note here that it's not sure that wether high-Eddington accreting systems in this paper have a similar broad line region size vs. 5100\,\AA\,luminosity relation, thus cause uncertainty of black hole mass estimation, this paper will not talk about this uncertainty. Among 25 objects in our final sample, 17 ($\sim 68\%$) objects are at the super-Eddington regime, whereas the rest of 8 (32\%) are high-Eddington sources, including four objects having lower limits. For this reason, the super-Eddington accreting AGN candidates used in this work will be referred to as extremely high Eddington ratio accreting supermassive black holes (EESBHs) with an Eddington ratio close to or above 1.

\section{VLA data reduction} \label{sec:data}

All the relevant raw data (visibilities) have been retrieved from the NRAO Data Archive\footnote{\url{https://archive.nrao.edu/archive/advquery.jsp}}, and only the data with good quality and the correct observational configurations have been adopted. In general, each target should have been observed for more than 20 seconds and followed by one exposure of a nearby phase calibrator. Although some results have already been published (see Table \ref{tab:vla}), in order to ensure uniformity of analysis across all the parameters, we performed a manual calibration for all the datasets using the Common Astronomy Software Application \citep[CASA v5.1.1,][]{2007ASPC..376..127M}. The archive data have a time range from 1980 to 2016 and are composed of two distinct groups, the historical or ordinary VLA products and the Karl G. Jansky VLA (JVLA) products. The historical VLA datasets were scheduled as a single channel in only one or two spectral windows (SPWs), while the JVLA continuum observations are all in the multi-SPW mode with each SPW having a good bandwidth coverage.

Our data analysis followed the standard routines described in CASA Cookbook. For the historical VLA datasets, we adopted a closest flux density standard concerning the observing date to get the absolute flux density for the primary flux calibrator, and subsequently bootstrapped onto the secondary flux density calibrators and targets. As an example, we use the flux density standards `Perley-Butler 2013' \citep{2013ApJS..204...19P} to set the overall flux density scale for datasets from the project 15A-283 running from July to September 2015. We also determined the gain solutions by using a nearby secondary calibrator and transferred them to target sources. In contrast to using a narrow band in historical VLA datasets, the JVLA datasets were all scheduled with multiple channels and spectral windows. We thus referred to the calibrating scheme/routine as described in most recent CASA Cookbook. In addition to transferring the flux density scale and gain solutions, antenna delay and bandpass corrections were also determined by fringe-fitting the visibilities.

Deconvolution and self-calibration were performed in DIFMAP \citep{1994BAAS...26..987S}. Only for sources with a high signal-to-noise ratio (SNR$>9$), self-calibration was applied using a well-established model. The self-calibration was performed initially only on phase, and subsequently on both phase and amplitude when we achieved a good model. Finally, the natural-weighting images were used for model fitting and further discussion. Two-dimensional Gaussian model was used to fit the visibility data of each target to obtain characteristic parameters, such as the integrated and peak flux density, as well as the full width at half maximum (FWHM) of the Gaussian model. The model-fitting results are listed in Table \ref{tab:results}. We also list some special data reduction notes in Appendix \ref{appendix}.

We estimated the uncertainty of the integrated flux density $S_i$ based on the formulae given by \citet{2003AJ....125..465H}. As our sources have an unresolved point-like nucleus, the relative errors of the integrated flux density $S_i$ were estimated from $\sigma_i/S_i =\sqrt{2.5\sigma_{rms}^2/S_p^2+{0.01}^2}$ \citep[][Eq.5]{2003AJ....125..465H}. The uncertainty of the peak flux density was estimated by combining the model fitting errors and the initial calibrating errors, which is $\sigma_p=\sqrt{\sigma_{rms}^2+{(1.5\sigma_{rms})}^2} \approx 1.8\sigma_{rms}$ \citep[][see their Appendix B]{2012AJ....144..105H}, where $1.5\sigma_{rms}$ comes from CLEAN error and $\sigma_{rms}$ is the experimental error.

\section{Results} \label{sec:results1}

The new results for 25 EESBHs observed with the VLA are obtained from a total of 45 datasets, of which 26 were previously published (see the references in Column 9 of Table \ref{tab:vla}), and the rest of 19 datasets are analyzed in this work. All but one object (IRAS 04416+1215) has the C-band data, and 22 out of 24 objects are detected at C-band with a peak flux density above $3\sigma$. All the sources were observed with a resolution of $<2$\,kpc at C-band, while 20 out of 24 sources were observed with a resolution of $<1$\,kpc. Table \ref{tab:results} lists the $5$\,GHz luminosity (column 8), the radio brightness temperature (column 9), and radio loudness (column 10).

The radio brightness temperature was estimated by using the formula \citep[e.g.][]{2005ApJ...621..123U}: 
\begin{equation}\label{eq:bt}
T_\mathrm{B}=1.8\times10^9(1+z)\frac{S_i}{\nu^2\theta^2}~\mathrm{(K)},
\end{equation}
where \(S_i\) is the integrated flux density of each Gaussian model component in units of mJy (column 5 of Table \ref{tab:results}); $\theta$ is the $\mathrm{FWHM}$ of the Gaussian model in milli-arcsec (column 7 of Table \ref{tab:results}); \(\nu\) is the observing frequency in GHz (column 2 of Table \ref{tab:results}), and \(z\) is the redshift. The estimated 5\,GHz and 8.4\,GHz radio brightness temperature are listed in Column 9 of Table \ref{tab:results}. Because the measured component size is just the upper limit, the radio brightness temperature should be considered as a lower limit. The non-simultaneous $5 - 8$\,GHz spectral index $\alpha$ (defined as $S_i \propto \nu^\alpha$) based on the integrated flux densities are listed in Column 11 of Table \ref{tab:results}.

\begin{figure}
\centering
\includegraphics[scale=1.35]{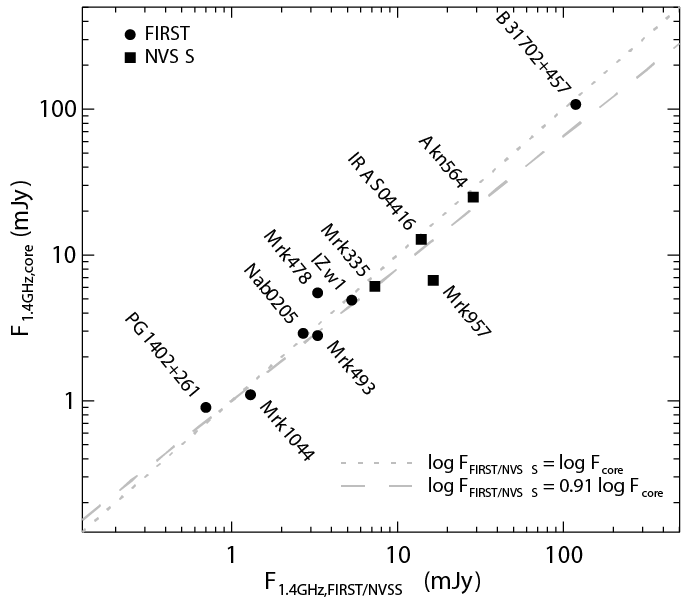}
\caption{A comparison between the 1.4\,GHz radio emission observed in VLA A-array and VLA B/D-array. The dotted and dashed line 
indicate $\log F_\mathrm{FIRST/NVSS}=\log F_\mathrm{core}$ and $\log F_\mathrm{FIRST/NVSS}=0.91\log F_\mathrm{core}$, respectively. 
} \label{fig:1.4}
\end{figure}

Next, we compared the radio emission at different galactic scales. For instance, the VLA A-array flux density at L-band may be taken as a measurement of the core radio emission, and the flux density derived from the Faint Images of the Radio Sky at Twenty-Centimeters (FIRST) 1.4 GHz survey \citep{1995ApJ...450..559B} made with the VLA B-array may serve as a proxy of the radio emission from the entire galaxy. These two observations provide a projected spatial resolution of $\sim4-8$\,kpc and $\sim20-30$\,kpc, respectively. The typical size of the optical hosts of EESBHs in our sample is comparable with the beam size of VLA B-array at 1.4\,GHz (the FIRST images). For those sources without the FIRST survey coverage or without detection, the NRAO VLA Sky Survey \citep[NVSS, ][]{1998AJ....115.1693C} flux density has been taken instead, and the 1.4\,GHz flux densities are listed in Table \ref{tab:1.4}, there are 11 sources in total. Figure \ref{fig:1.4} shows the comparison between the FIRST/NVSS flux density and the VLA A-array 1.4\,GHz flux density. Taking a 1\,$\sigma$ uncertainty, the two fluxes are in good agreement for the majority of sources. Nine out of 11 sources have a central flux density that accounts for $>$80\% of the total radio emission. Among these sources, the average flux density difference between VLA A-array and FIRST/NVSS measurements is only 18\% (Mrk\,957 has a 59\% VLA A-array flux density decrease by comparing it with the NVSS measurement). We have also marked a line $\log F_\mathrm{FIRST/NVSS}=0.91\log F_{core}$ in Figure \ref{fig:1.4} to show the comparison between the FIRST/NVSS and the VLA A-array 1.4\,GHz flux density. This indicates that most sources are compact at this resolution and the radio emission is dominated by the central $\sim$4-8\,kpc region. Furthermore, as the time lag between FIRST, NVSS and the VLA A-array 1.4\,GHz observations, we can infer that our EESBHs have a quite stable radio emission.


\begin{table*}
\small
\renewcommand\arraystretch{0.8}                                                                                                                                                                              
\centering                                                                                                                                                                                 
\caption{The observational results and model fitting parameters}\label{tab:results}                                                                                                                                                                                                                                                                                                                                                   
\begin{tabular}{rrrrrrrrrrrc}                                                                                                                                                         
\toprule 

\tc{Short Name}  &  \tc{Freq.}   &  \tc{Res.}    &  \tc{$S_p$}                       &   \tc{$S_i$}  &  \tc{rms}              &  \tc{$\theta$}  &  \tc{log\,$L_5$}          & \tc{log\,$T_\mathrm{B}$}  & \tc{$\mathcal{R}$}  & \tc{$\alpha_{5}^{8.4}$}    &  \tc{log\,$\Sigma_\mathrm{SFR}$}                  \\
                           &  \tc{(GHz)}  &  \tc{(kpc)}    &  \tc{(mJy\,beam$^{-1}$)}  &   \tc{(mJy)}  &  \tc{($\mu$Jy)}    &  \tc{(mas)}     &   \tc{(erg\,s$^{-1}$)}  &   \tc{(K)}                          &                                &       & ($*$)     \\                                                                                                                                     
\tc{(1)}   &  \tc{(2)}   &   \tc{(3)}    &          \tc{(4)}                 &             \tc{(5)}          &        \tc{(6)}        &         \tc{(7)}      &       \tc{(8)}              &       \tc{(9)}          &       \tc{(10)}                   &    \tc{(11)}      &    \tc{(12)}                  \\                                                                                                                                     

\hline                                                                                                                                                          
J0230-0859           &   4.86  &$  0.13  $&$   0.44  \pm  0.11          $&$     0.44  \pm  0.09    $&$      61.74      $&$       55.12    $&$     37.12           $&$      4.05        $&$    0.22      $&$       -0.73   $&$    1.02 $\\
                     &   1.48  &$  0.42  $&$   1.09  \pm  0.37          $&$     1.10  \pm  0.32    $&$     207.25      $&$                $&$                     $&$                  $&$              $&$               $&$   -0.02\;\;\, $\\
                     &   8.46  &$  0.07  $&$   0.28  \pm  0.07          $&$     0.29  \pm  0.07    $&$      43.90      $&$       68.89    $&$                     $&$      3.20        $&$              $&$               $&$    1.46 $\\
J2242+2943           &   5.17  &$  0.22  $&$   5.69  \pm  0.03          $&$     5.84  \pm  0.03    $&$      21.11      $&$      161.50    $&$     38.63           $&$      4.18        $&$    4.02      $&$       -1.23   $&$    2.33 $\\
                     &   1.42  &$  0.71  $&$  22.36  \pm  0.85          $&$    24.92  \pm  0.84    $&$     477.40      $&$      499.21    $&$                     $&$                  $&$              $&$               $&$    1.48 $\\
                     &   8.46  &$  0.10  $&$   2.81  \pm  0.07          $&$     3.18  \pm  0.07    $&$      40.87      $&$       94.40    $&$                     $&$      3.96        $&$              $&$               $&$    2.83 $\\
J1153+4612           &   4.86  &$  0.20  $&$   0.21  \pm  0.09          $&$     0.22  \pm  0.08    $&$      54.19      $&$      209.30    $&$     37.22           $&$      2.61        $&$    0.75      $&$               $&$    0.88 $\\
J0006+2012           &   5.17  &$  0.20  $&$   3.04  \pm  0.03          $&$     3.21  \pm  0.02    $&$      17.06      $&$       92.61    $&$     38.44           $&$      4.40        $&$    1.05      $&$       -0.84   $&$    2.13 $\\
                     &   1.42  &$  0.80  $&$   6.15  \pm  0.33          $&$     6.19  \pm  0.29    $&$     184.79      $&$      277.85    $&$                     $&$                  $&$              $&$               $&$    0.79 $\\
                     &   8.43  &$  0.12  $&$   1.92  \pm  0.07          $&$     2.14  \pm  0.06    $&$      39.25      $&$       60.59    $&$                     $&$      4.18        $&$              $&$               $&$    2.58 $\\
J1559+3501           &   5.17  &$  0.21  $&$   0.48  \pm  0.02          $&$     0.62  \pm  0.03    $&$      15.64      $&$      286.15    $&$     37.87           $&$      2.72        $&$    1.20      $&$       -0.17   $&$    1.00 $\\
                     &   1.40  &$  0.90  $&$   2.11  \pm  0.33          $&$     2.85  \pm  0.39    $&$     183.58      $&$     1089.62    $&$                     $&$                  $&$              $&$               $&$    0.54 $\\
                     &   8.84  &$  0.16  $&$   0.56  \pm  0.01          $&$     0.56  \pm  0.01    $&$       4.56      $&$       18.21    $&$                     $&$      4.60        $&$              $&$               $&$    2.13 $\\
J1703+4540           &   5.10  &$  0.46  $&$  26.39  \pm  0.03          $&$    27.50  \pm  0.03    $&$      19.01      $&$       92.85    $&$     40.10           $&$      5.36        $&$   58.00      $&$       -0.56   $&$    2.88 $\\
                     &   1.42  &$  1.65  $&$  99.26  \pm  0.36          $&$   107.89  \pm  0.34    $&$     200.73      $&$      276.37    $&$                     $&$                  $&$              $&$               $&$    2.14 $\\
                     &   8.43  &$  0.27  $&$  20.21  \pm  0.15          $&$    20.73  \pm  0.13    $&$      85.34      $&$       49.39    $&$                     $&$      5.35        $&$              $&$               $&$    3.59 $\\
J1536+5433           &   5.17  &$  0.29  $&$   0.64  \pm  0.02          $&$     0.65  \pm  0.02    $&$      15.36      $&$       18.93    $&$     38.10           $&$      5.10        $&$    0.81      $&$               $&$    1.29 $\\
J1034+3938           &   5.17  &$  0.34  $&$   6.75  \pm  0.01          $&$     6.98  \pm  0.01    $&$      10.06      $&$       78.36    $&$     39.19           $&$      4.90        $&$   12.18      $&$               $&$    2.44 $\\
J1242+3317           &   5.17  &$  0.31  $&$   1.53  \pm  0.02          $&$     1.37  \pm  0.02    $&$      14.58      $&$      127.77    $&$     38.51           $&$      3.77        $&$    3.87      $&$               $&$    1.80 $\\
J1025+5140           &   4.86  &$  0.37  $&$   0.19  \pm  0.07          $&$     0.22  \pm  0.07    $&$      40.03      $&$      117.40    $&$     37.73           $&$      3.10        $&$    0.29      $&$               $&$    0.92 $\\
J1431+2817           &   4.86  &$  0.39  $&$\sim0.05  \pm  0.02         $&$ \sim0.05  \pm  0.02    $&$      15.78      $&$  \sim156.97    $&$ \sim37.09           $&$  \sim2.20        $&$\sim0.03      $&$               $&$    0.23 $\\
J1246+0222           &   5.17  &$  0.39  $&$   0.92  \pm  0.04          $&$     0.85  \pm  0.03    $&$      22.28      $&$       80.06    $&$     38.40           $&$      3.97        $&$    1.38      $&$               $&$    1.48 $\\
J0444+1221           &   5.00  &$        $&$                            $&$         \sim4.07       $&$                 $&$                $&$ \sim39.62           $&$                  $&$\sim4.88      $&$               $&$         $\\
                     &   1.42  &$  2.19  $&$  12.14  \pm  0.28          $&$    12.83  \pm  0.26    $&$     158.73      $&$      317.15    $&$                     $&$                  $&$              $&$               $&$    1.32 $\\
                     &   8.43  &$  0.36  $&$   2.04  \pm  0.12          $&$     2.52  \pm  0.13    $&$      70.82      $&$      113.48    $&$                     $&$      3.73        $&$              $&$               $&$    2.79 $\\
J1014-0418           &   4.86  &$  0.44  $&$   0.47  \pm  0.13          $&$     0.48  \pm  0.11    $&$      73.44      $&$       62.94    $&$     38.30           $&$      3.99        $&$    0.19      $&$               $&$    1.16 $\\
J0053+1241           &   4.86  &$  0.45  $&$   1.85  \pm  0.11          $&$     1.89  \pm  0.10    $&$      64.90      $&$       61.95    $&$     38.90           $&$      4.59        $&$    0.37      $&$       -1.11   $&$    1.91 $\\  
                     &   1.42  &$  1.55  $&$   4.63  \pm  0.29          $&$     4.91  \pm  0.27    $&$     161.83      $&$      291.11    $&$                     $&$                  $&$              $&$               $&$    0.83 $\\
                     &   8.43  &$  0.28  $&$   0.82  \pm  0.02          $&$     1.02  \pm  0.03    $&$      15.70      $&$      122.42    $&$                     $&$      3.26        $&$              $&$               $&$    2.28 $\\
J1451+2709           &   4.86  &$  0.52  $&$   1.37  \pm  0.09          $&$     1.43  \pm  0.09    $&$      55.42      $&$      86.13     $&$     38.87           $&$      4.19        $&$    0.81      $&$               $&$    1.77 $\\
J0041+4021           &   4.86  &$  0.53  $&$   1.61  \pm  0.08          $&$     2.08  \pm  0.09    $&$      44.74      $&$      246.43    $&$     39.11           $&$      3.44        $&$    9.56      $&$               $&$    1.95 $\\             
                     &   1.42  &$  1.95  $&$   5.82  \pm  0.21          $&$     6.78  \pm  0.22    $&$     119.44      $&$  \sim400.00    $&$                     $&$                  $&$              $&$               $&$    0.99 $\\
J0107+1408           &   4.86  &$  0.56  $&$  <0.06  \pm  0.04          $&$    <0.06  \pm  0.03    $&$      22.28      $&$                $&$    <37.67           $&$                  $&$   <1.13      $&$               $&$    0.51 $\\   
J1442+3526           &   4.86  &$  0.63  $&$   0.96  \pm  0.10          $&$     1.83  \pm  0.17    $&$      59.15      $&$      510.15    $&$     39.13           $&$      2.76        $&$    0.73      $&$       -2.09   $&$    1.91 $\\       
                     &   1.42  &$  1.74  $&$   3.65  \pm  0.31          $&$     5.55  \pm  0.42    $&$     176.10      $&$      842.75    $&$                     $&$                  $&$              $&$               $&$    1.06 $\\     
                     &   8.46  &$  0.34  $&$   0.31  \pm  0.11          $&$     0.57  \pm  0.18    $&$      63.17      $&$      219.44    $&$                     $&$      2.50        $&$              $&$               $&$    2.11 $\\
J1140+0307           &   4.86  &$  0.61  $&$\sim0.05 \pm  0.02          $&$ \sim0.06  \pm  0.03    $&$      16.13      $&$  \sim214.29    $&$ \sim37.68           $&$  \sim2.02        $&$\sim0.76      $&$               $&$    0.36 $\\
J1355+5612           &   5.17  &$  0.86  $&$   2.11  \pm  0.05          $&$     2.23  \pm  0.04    $&$      29.09      $&$      128.06    $&$     39.67           $&$      4.00        $&$    7.30      $&$               $&$    1.70 $\\
J1118+4025           &   4.86  &$  1.13  $&$  <0.16  \pm  0.09          $&$    <0.16  \pm  0.08    $&$      54.23      $&$                $&$    <38.72           $&$                  $&$   <0.86      $&$               $&$    0.97 $\\
J0207+0242           &   4.86  &$  1.11  $&$   1.17  \pm  0.10          $&$     1.23  \pm  0.10    $&$      60.94      $&$      133.26    $&$     39.61           $&$      3.78        $&$    1.04      $&$               $&$    1.71 $\\
                     &   1.46  &$  3.84  $&$   2.85  \pm  0.29          $&$     2.95  \pm  0.27    $&$     166.59      $&$      311.50    $&$                     $&$                  $&$              $&$               $&$    0.69 $\\
J1405+2555           &   4.86  &$  1.13  $&$   0.26  \pm  0.05          $&$     0.27  \pm  0.04    $&$      31.20      $&$       93.90    $&$     39.00           $&$      3.43        $&$    1.13      $&$               $&$    1.25 $\\
                     &   1.42  &$  3.89  $&$   0.93  \pm  0.29          $&$     0.95  \pm  0.26    $&$     163.47      $&$       77.54    $&$                     $&$                  $&$              $&$               $&$    0.29 $\\
J2236+1343           &   4.86  &$  1.94  $&$   0.24  \pm  0.09          $&$     0.24  \pm  0.07    $&$      50.71      $&$      394.22    $&$     39.61           $&$      2.19        $&$    0.86      $&$        1.07   $&$    1.32 $\\
                     &   1.47  &$  6.22  $&$   0.92  \pm  0.30          $&$     0.90  \pm  0.26    $&$     171.49      $&$      169.46    $&$                     $&$                  $&$              $&$               $&$    0.51 $\\
                     &   8.46  &$  1.21  $&$   0.42  \pm  0.03          $&$     0.43  \pm  0.02    $&$      18.19      $&$       62.41    $&$                     $&$      3.57        $&$              $&$               $&$    2.25 $\\

\hline                                                                                                                                                                                                                                                
\end{tabular}

\begin{flushleft}                                                                                                                                                                                                                                     
\textbf{Notes}: Column 1: short name; Column 2: central frequency; Column 3: physical resolution                                                                                                                                     
in kpc; Column 4: peak flux density, '$\sim$' means this source is only marginally detected, and for                                                                                                                                     
the sources have not detected we take $3\sigma$ upper limit; Column 5: integrated flux. The C-band                                                                                                                                  
flux density for source J0444+1221 is estimated from X-band data; Column 6: image noise root-mean-square;                                                                                                                  
Column 7: $FWHM$ of Gaussian model;                                                                                                                                                                                            
Column 8: 5\,GHz radio luminosity; Column 9: the lower limit of radio brightness                                                                                                                                     
temperature; Column 10: radio loudness; Column 11: non-simultaneous 5 to 8.4\,GHz spectral index;                                                                                                                             
Column 12: lower limit of SFR surface density ($*$: $\mathrm{M_\odot\,yr^{-1}\,kpc^{-2}}$) estimated from radio luminosity. \\                                                                                                                                                                                                                                                       
\end{flushleft}

\end{table*}                                                                                                                                                                                                                                          

\section{Parameter correlations} \label{sec:results2}

The correlation between the radio loudness $\mathcal{R}$ and the Eddington ratio $\lambda_\mathrm{Edd}$ in the extremely high-Eddington regime ($\lambda_\mathrm{Edd}\gtrsim1$) has not been well explored previously because of the limited number of super-Eddington sources and a poor estimation of their Eddington ratios. In order to form a broad parameter space for the Eddington ratio, we include 199 AGNs from \citet{2007ApJ...658..815S} having a wide range of Eddington ratios from log\,$\lambda_\mathrm{Edd}=-7$ to $1$. 

An accurate estimate of the bolometric luminosity is required to get the Eddington ratio $\lambda_\mathrm{Edd}$, which is defined as $\lambda_\mathrm{Edd} \equiv L_\mathrm{bol}/L_\mathrm{Edd}$. The integral of the observed SED of the optical-ultraviolet and X-ray emission by the accretion disc and the hot corona provides a direct measurement of the bolometric luminosity $L_\mathrm{bol}$. One generally uses the $B$-band continuum luminosity $L_\mathrm{B}$ to estimate the bolometric luminosity with a bolometric correction factor $\kappa_\mathrm{B}$ expressed as $L_\mathrm{bol}=\kappa_\mathrm{B} L_\mathrm{B}$. The commonly used $B$-band bolometric correction factor is $\kappa_\mathrm{B}=10$ \citep[e.g.][]{2004MNRAS.351..169M, 2004MNRAS.352.1390M}. However, it is only suitable for sub-Eddington systems as the high-Eddington ratio sources tend to have a larger bolometric correction factor \citep[see][]{2007MNRAS.381.1235V}. 

It was also noted by \citet{2012MNRAS.425..907J, 2012MNRAS.420.1825J} that NLS1s often possess high Eddington ratios, having significantly higher values of the optical bolometric correction factor than other samples. Surprisingly, the average value of $5100$\,\AA\,bolometric correction factor $\kappa_{5100}\approx70$ has been found for 10 NLS1s. \citet{2013PhRvL.110h1301W} independently got $\kappa_{5100}\approx 40 \sim 100$ for their super-Eddington accreting AGN candidates. The majority of high-Eddington sources in \citet{2007ApJ...658..815S} are Palomar-Green (PG) quasars; therefore, instead of using a common bolometric correction factor, we use the bona fide bolometric luminosity which was obtained from the SED fitting to re-calculate the Eddington ratios for PG quasars and our EESBHs. The bolometric luminosities for our EESBHs and PG quasars are presented in Table \ref{tab:sample} (column 7) and Table \ref{tab:pg}, respectively. On the other hand, the $B$-band bolometric correction factor $\kappa_B=10$ was used to set a lower limit of the Eddington ratios for those EESBHs without previous bolometric luminosity estimates.

The radio loudness of these 199 sources are taken from \citet{2011MNRAS.417..184B}, measured base on the core-only radio emission. The strategy of using the core-only radio emission depends on the idea that the optical/X-ray radiation is only related to the current/recent nuclear activity, which is not directly correlated with large scale radio emission. It was found that the jet emission can exist for a timescale of $\sim10^{7-8}$ years \citep[e.g.,][]{2018arXiv180909008K}. On the other hand, the large-scale jet emission is not only dependent on the central engine itself but also the surrounding environment. Such as radio galaxies as well as radio-loud quasars reside in denser environments than radio-quiet AGNs \citep[e.g.][]{2011A&A...535A..21L}. Besides, the star formation in host galaxies will also contribute significant radio emission on the whole galaxy scale. Since the radio loudness and the Eddington ratio is relatively dependent on the optical or X-ray estimates, it is reasonable to adopt the core-only radio emission for the relation between radio loudness and the Eddington ratio.

\begin{table*}                                                                                      
\renewcommand\arraystretch{0.8}                                                                                                                                                                              
\centering                                                                                                                                                                                 
\caption{VLA observational results at L-band (1.4\,GHz)} \label{tab:1.4}                                                                                                                                                                       
\label{tab:lband}                                                                                                                                                                            
\begin{tabular}{lllllll}                                                                                                                                                         
\toprule 

Name     &   Mrk\,1044     &    Akn\,564        &      Mrk\,335       &     Mrk\,493        &     B3\,1702+457     &      IRAS\,04416+1215     \\
\hline
VLA-B/D  &$  1.3\pm0.1^F  $&$   28.6\pm0.9^N   $&$     7.3\pm0.5^N   $&$    3.3\pm0.1^F    $&$    118.6\pm0.1^F   $&$     13.9\pm0.6^N    $\\
VLA-A    &$  1.1\pm0.2    $&$   24.9\pm0.4     $&$     6.1\pm0.1     $&$    2.8\pm0.1      $&$    107.8\pm0.2     $&$     12.8\pm0.1      $\\
\hline                                                                                        
Name     &   IZw1          &    Mrk\,957        &      Mrk\,478       &     Nab\,0205+024   &     PG\,1402+261     &                        \\
\hline                                                                                                               
VLA-B/D  &$  5.3\pm0.1^F  $&$   16.4\pm0.6^N   $&$     3.3\pm0.1^F   $&$    2.7\pm0.1^F    $&$    0.7\pm0.1^F     $&$                     $\\
VLA-A    &$  4.9\pm0.1    $&$  \;\;6.7\pm0.1   $&$     5.5\pm0.1     $&$    2.9\pm0.1      $&$    0.9\pm0.1       $&$                     $\\

\hline                                                                                                                                                                                            
\end{tabular}                                                                                                                                                                                   
                                                                                                                                                                                                                                                                                                                                                                                                                          
\textbf{Notes}: 1.4\,GHz flux density in mJy. $^N$: NVSS; $^F$: FIRST                                                                                                                                                                                                
                                                                                                                                                                                                
\begin{flushleft}                                                                                                                                                                               
\end{flushleft}

\end{table*}                                                                                                                                                                                    

\begin{figure}
\centering
\includegraphics[scale=0.32]{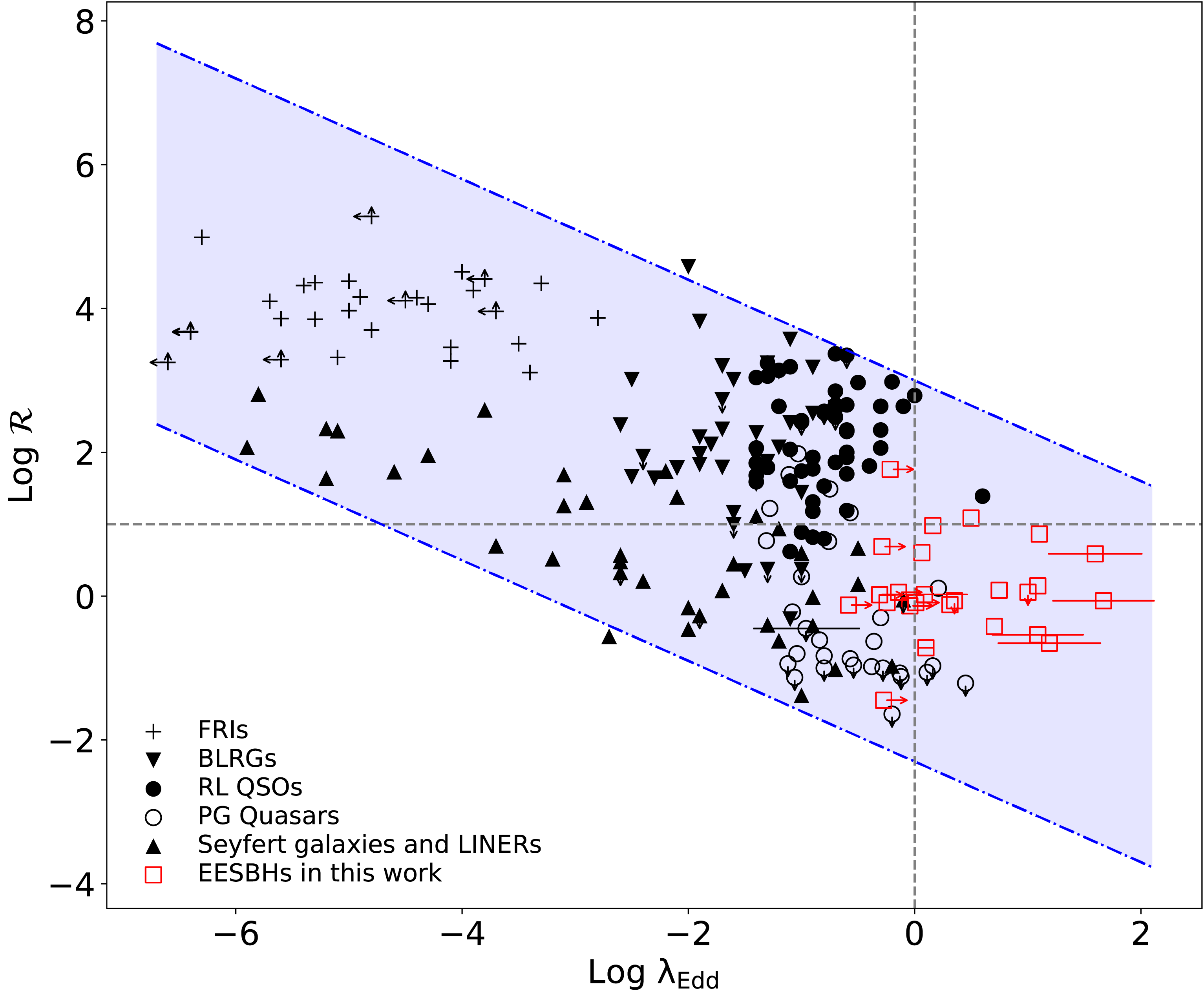}
\caption{Radio loudness $\mathcal{R}$ vs Eddington ratio $\lambda_\mathrm{Edd}$. The markers are designated in the left-bottom corner and the error bars in some EESBH objects come from the uncertainty of the BH spin. The vertical dotted line is $\mathrm{\lambda_{Edd}}$=1 and the horizontal dotted line at $\mathcal{R}=10$ represents the division between radio-loud (above) and radio-quiet (below) sources. The blue belt shows the upper and lower boundaries corresponding to the slope of $-0.73$.}\label{fig:re}
\end{figure}

The standard radio loudness parameter \citep{1989AJ.....98.1195K} is defined as $\mathcal{R}\equiv L_{\nu_5}/L_{\nu_B}=1.3\times10^5(L_{5}/L_\mathrm{B})$, where the $L_5$ is the radio luminosity at 5\,GHz and $L_\mathrm{B}$ is the optical luminosity of the nucleus at $\lambda_B = 4400$\,\AA; both are measured in erg s$^{-1}$. For our EESBHs, we have measured the core radio emission from the brightest component of the object. Nearly all of our EESBHs are unresolved with VLA A-array observation at 5\,GHz (except for Akn\,564), and most have a component size of a few hundred parsecs (except for four sources with relatively large redshift, i.e. $z>0.15$). Therefore, the 5\,GHz core-only radio emission of our EESBHs is consistent with \citet{2011MNRAS.417..184B}. Here the total radio luminosity is expressed as $L_5 \equiv \nu_5 L_{\nu_5}$ and the total $B$-band luminosity as $L_\mathrm{B} \equiv \nu_B L_{\nu_B}$. In this paper, we calculate the $B$-band luminosity by using the 5100\,\AA\,luminosity using the transformation $L_\mathrm{B}=L_{5100}\times(\lambda_B/5100)^{1+\alpha_{opt}}$ and assuming a constant optical spectral index $\alpha_{opt}=-0.5$ \citep[e.g.][]{2007ApJ...658..815S}. The radio luminosity is given as: 
\begin{equation}
L_\mathrm{R}=\nu_R f_R \frac{4\pi D_L^2}{(1+z)^{1+\alpha_R}}
\end{equation}
where $\alpha_R$ is the radio spectral index, assuming $\alpha_R=-0.8$ according to our measurement of $5-8.4$\,GHz spectral index, $f_R$ is the radio flux density, and $D_L$ is the luminosity distance. Figure \ref{fig:re} shows the radio loudness versus the Eddington ratio for 199 comparison AGNs and our EESBHs. Here we keep the traditional labels of the source sample for 199 comparison AGNs, similarly hereinafter. The linear regression slope between $\mathrm{log}\mathcal{R}$ and $\mathrm{log}\lambda_\mathrm{Edd}$ is $-0.73\pm0.08$ within the $95\%$ confidence interval, the blue belt shows the upper and lower boundaries corresponding to the slope of $-0.73$. Interestingly, we can see that there are a lot of vacancy both at upper left and lower right in Figure \ref{fig:re}, which imply that our sample is incomplete for larger radio loudness at the lower boundary of the Eddington ratios, and also for both larger and lower radio loudness at the higher boundary of Eddington ratios. However, the trend may also hint a real features of radio loudness at higher and lower Eddington ratios.

\begin{figure}
\centering
\includegraphics[scale=0.32]{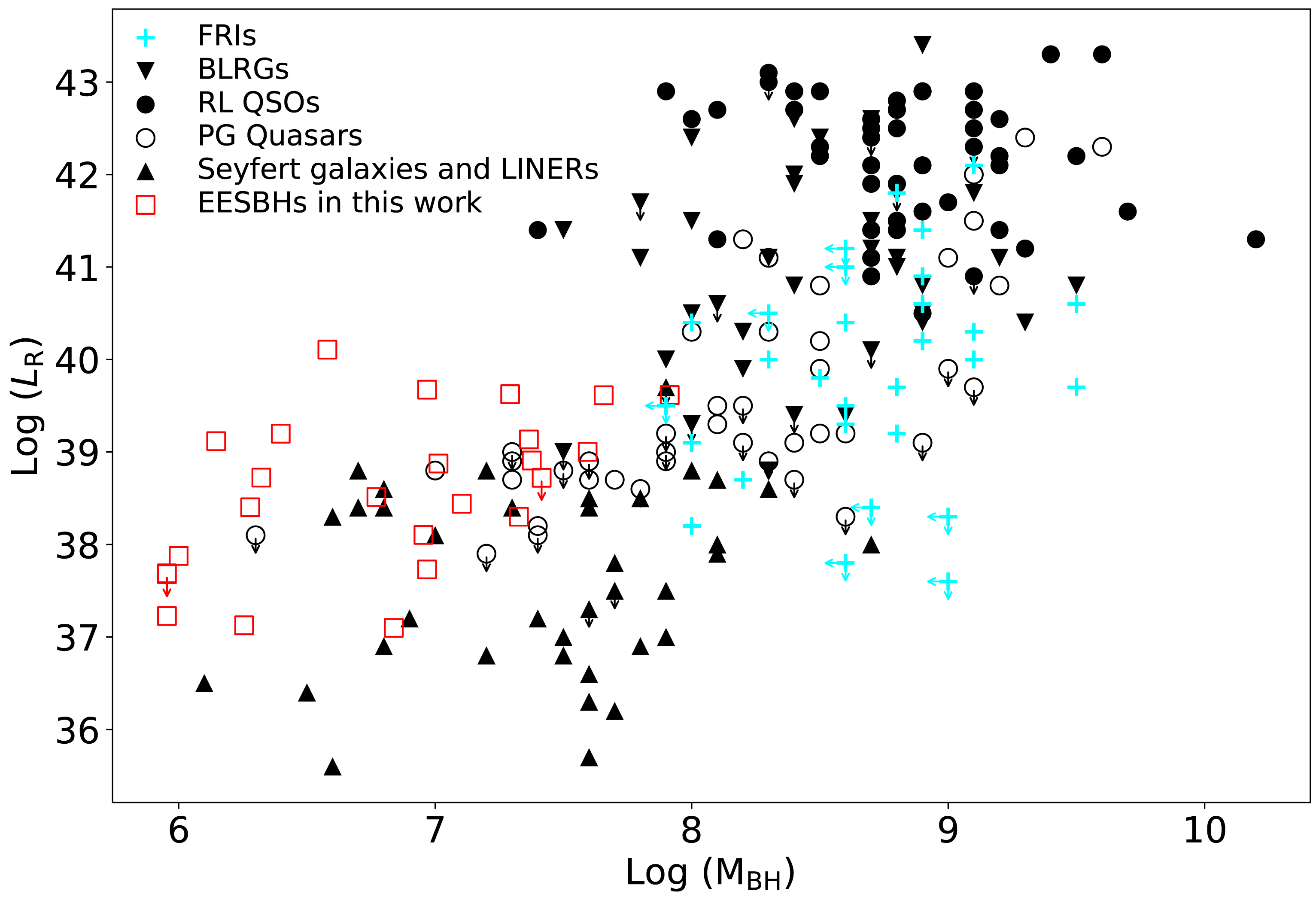}
\caption{The 5\,GHz core radio luminosity $L_\mathrm{R}$ vs. black hole mass $\mathrm{M}_\mathrm{BH}$. The markers are designated in the top-left corner.}
\label{fig:lrmbh}
\end{figure}

The other problem is the regression slope between $\mathrm{log}\mathcal{R}$ and $\mathrm{log}\lambda_{Edd}$ is close to $-1$ and possibly governed by the mutual dependence of $\mathcal{R}$ and $\lambda_{Edd}$ on optical luminosity, i.e. according to the definitions, $\mathcal{R}\propto L_\mathrm{B}^{-1}$ and $\lambda_\mathrm{Edd}\propto L_\mathrm{bol}$, and the trouble spot is we take the relation of $L_\mathrm{bol}=10L_\mathrm{B}$ for the samples in Figure \ref{fig:re} except for PG Quasars and EESBHs. However we find that $\mathcal{R}\cdot \lambda_\mathrm{Edd} = 1.3\times10^6(L_\mathrm{R}/L_\mathrm{Edd})$ (being free of the parameter $L_\mathrm{B}$), if there is no correlation between $L_\mathrm{R}$ and $L_\mathrm{Edd}$ or they are identical, then $\mathrm{log}\mathcal{R}$ vs. $\mathrm{log}\lambda_{Edd}$ naturally has a slope of $-1$ caused by the mutual dependence on $L_\mathrm{B}$. We have therefore explored the correlation between the 5\,GHz core radio luminosity $L_\mathrm{R}$ and the Eddington luminosity $L_\mathrm{Edd}$, here we use $\mathrm{M}_\mathrm{BH}$ to replace $L_\mathrm{Edd}$ due to the equation $L_\mathrm{Edd}=3.2\times10^4(M_\mathrm{BH}/M_\mathrm{\odot})L_\mathrm{\odot}$, see Figure \ref{fig:lrmbh}. There is a clear linear correlation between $L_\mathrm{R}$ and $\mathrm{M}_\mathrm{BH}$ and the linear regression gives a slope of $4.86\pm0.05$ at significance level of $>95\%$, which differs from $1$ and for that reason the correlation between $\mathrm{log}\mathcal{R}$ and $\mathrm{log}\lambda_{Edd}$ is not caused by the mutual dependence on $L_\mathrm{B}$.

\begin{figure}
\centering
\includegraphics[scale=0.33]{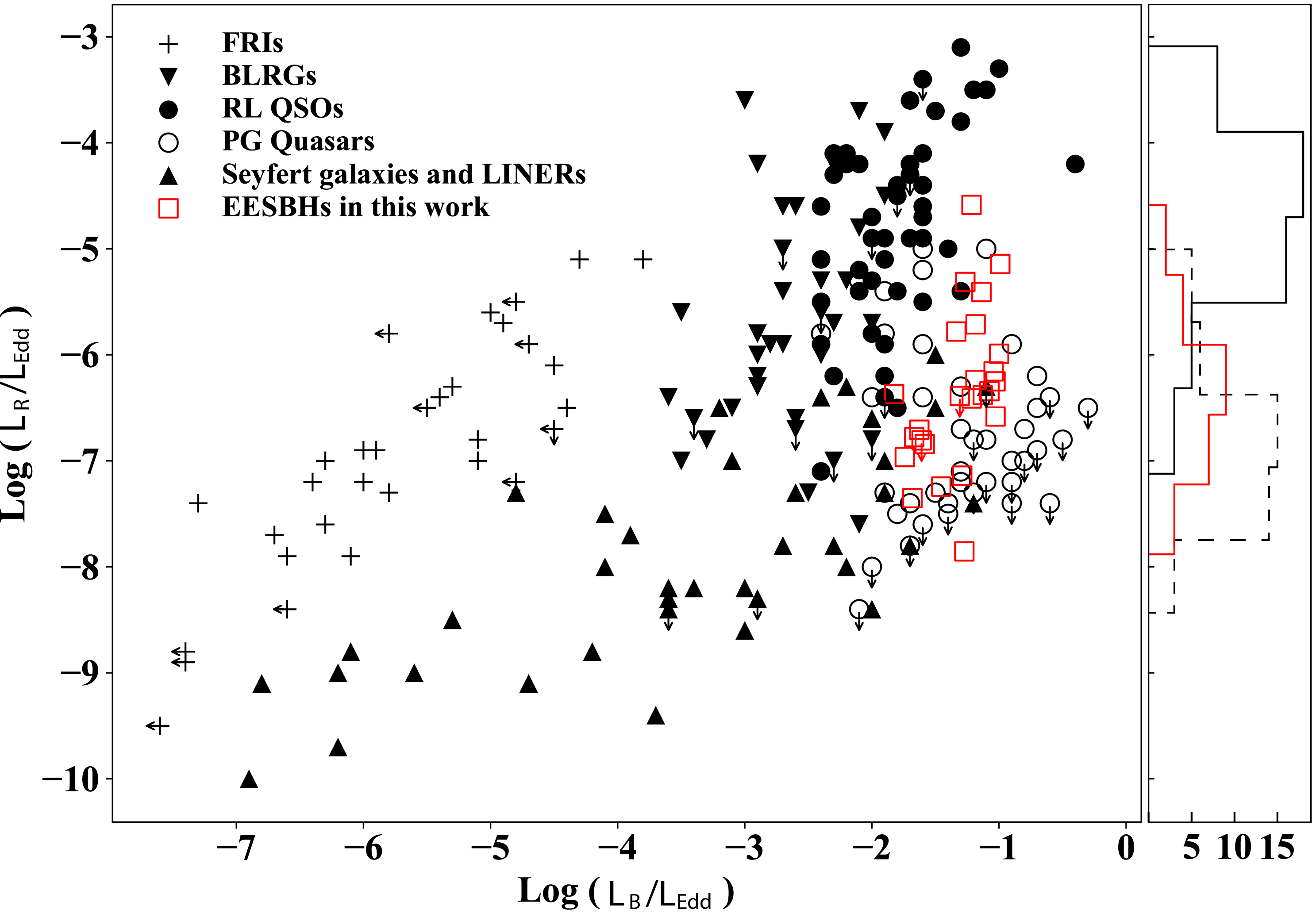}
\caption{The 5\,GHz core luminosity vs. the $B$-band luminosity, both in Eddington units. The markers are designated in the top-left corner. Right histogram: The black solid line histogram plots the distribution of $L_\mathrm{R}/L_\mathrm{Edd}$ for the RL QSOs, the black dashed line histogram is for PG Quasars and the solid red line is for EESBHs in this work.} \label{fig:ll}
\end{figure}

Figure \ref{fig:ll} shows the radio luminosity versus the optical luminosity expressed in Eddington units. There is no clear boundary between the sequence of radio-loud and radio-quiet sources in our plot in comparison with the results in \citet{2007ApJ...658..815S}. Our EESBHs located in the high end of optical luminosity and have an intermediate distribution of radio luminosity that lies between radio-loud quasars sample and Palomar-Green quasars sample (see the histogram in Figure \ref{fig:ll}).

\section{Discussion} \label{sec:dis}

\subsection{The origin of radio emission: AGN or star-forming activity?}

In the central kpc region of the radio-quiet AGN host galaxies, the AGN is often not the only source that can produce the observed radio emission. Recalling that the radio emission from galaxies may have various origins in Section \ref{sec:intro}, e.g. from stellar activities such as thermal free-free emission from HII regions \citep[e.g.,][]{1997ApJS..109..417L, 1997ApJ...488..621U}, non-thermal synchrotron radio emission \citep{2000A&A...354..423L} from young supernovae \citep[SNe, e.g.,][]{1985Sci...227...28K, 1992xrea.conf..247K, 2001ApJ...553L..19C, 2006ApJ...638..938A} and supernova remnants \citep[SNRs, e.g.,][]{1994MNRAS.266..455M, 1997ApJ...488..621U}, or from the jets of AGNs \citep[see][and references therein]{2008ARA&A..46..475H}. It's quite possible that the star-forming activities will dominate observed radio emission of the central kpc region in starburst galaxies \citep[e.g.][]{1997ApJ...488..621U, 2000A&A...358...95T, 2008MNRAS.391.1384F, 2011ApJ...740...95B}. Several observational characteristics can be used to discriminate between the radio emissions from AGNs and star-forming activities, such as the brightness temperature, the spectral index and the radio morphology \citep[see][and references therein]{2019NatAs...3..387P}. In the following sections, we explore the radio emission of our EESBHs by using these indicators.

\subsubsection{Radio morphology, spectral index and brightness temperature}

\begin{figure*}
\centering
\includegraphics[scale=0.4]{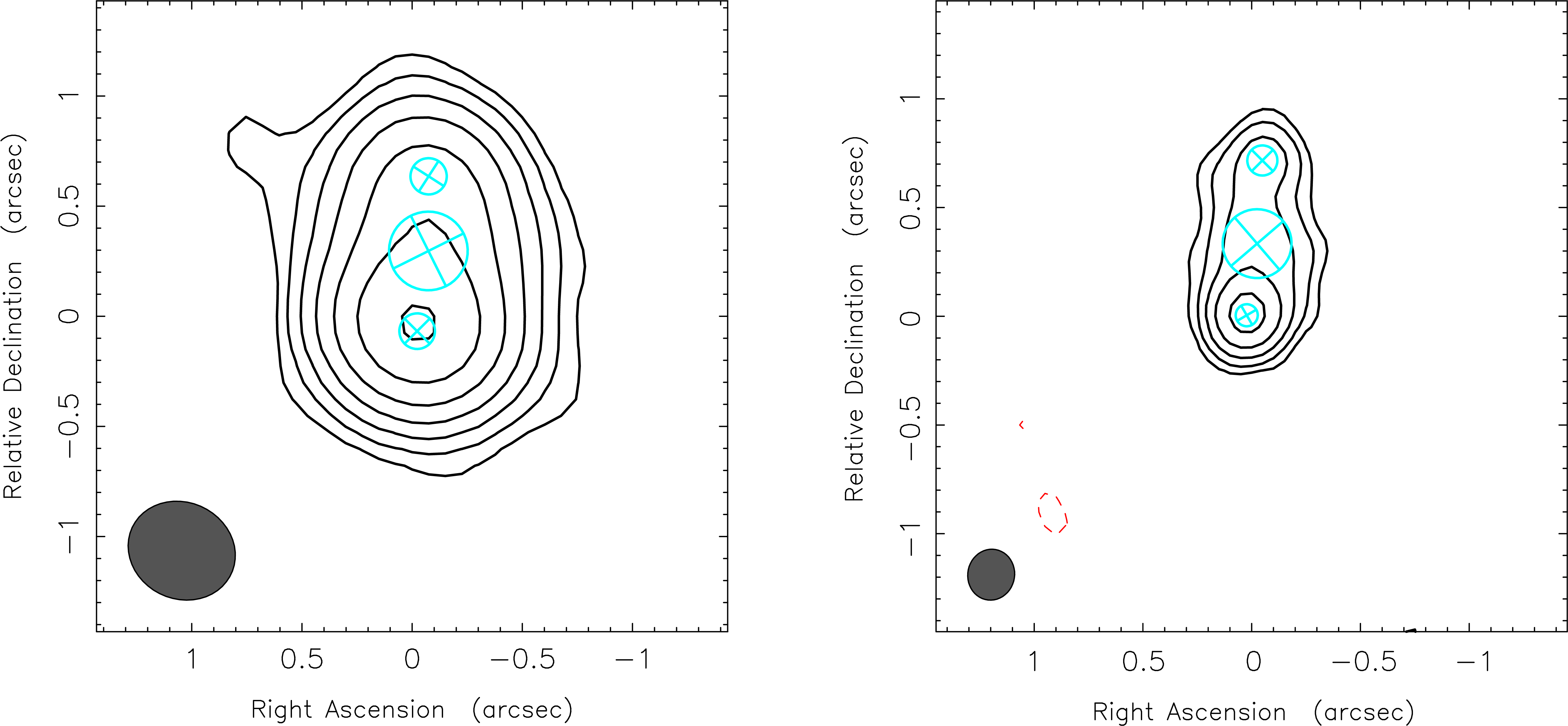}
\caption{The natural-weighting VLA A-array images of Akn\,564 at 5\,GHz (left) and 8.4\,GHz (right). Contours are drawn beginning at 3$\sigma$ and increase by a factor of 2 thereafter. The \textit{rms} noise is $\sigma_5=0.029$\,mJy\,beam$^{-1}$ and $\sigma_{8.5}=0.044$\,mJy\,beam$^{-1}$, respectively. The negative contours are plotted as red dashed lines and positive ones are plotted as the solid black lines. The restoring beam is shown in the left bottom of each panel. The \textit{FWHM} of Gaussian model components is plotted as cyan.}
\label{fig:akn564}
\end{figure*}

Previous studies show that an AGN is still the dominating radio-emitting source at radio flux densities above $0.1$\,mJy in radio-quiet AGNs \citep[][]{2011ApJ...740...20P, 2013MNRAS.436.3759B}. This indicates that the radio emission in most of our EESBHs could be from AGN activities and only three sources (Mrk\,684, SDSS\,J010712+140845 and SDSS\,J114008+030711) are below this threshold. In our EESBH sample, Akn\,564 is the only source showing evident AGN activity with a collimated linear structure extending $\sim1$\,kpc towards the north, and three resolved components in the VLA 5\,GHz and 8.4\,GHz images (see Figure \ref{fig:akn564}). Furthermore, two sources (PG\,2233+134 and Mrk\,493) show flat or inverted spectra, suggesting the radio emission is from AGNs.

Our EESBHs show a moderate brightness temperature range of $T_B\sim10^2$ - $10^5$\,K at both 5\,GHz and 8.4\,GHz as compared with radio-loud AGNs. Only two sources (B3\,1702+457 and Mrk\,486) have radio brightness temperatures above $10^5$\,K, i.e. $10^{5.36}$\,K and $10^{5.10}$\,K, respectively. Since a star-forming region cannot reach $T_B\sim10^5$\,K at frequency $\nu > 1$\,GHz \citep{1991ApJ...378...65C, 1992ASPC...31...79C, 1992ARA&A..30..575C}, we conclude that the radio emission observed in these two sources are preferentially from AGN activities.

Additionally, eight sources have a radio brightness temperature $T_B\gtrsim10^4$\,K (but $<10^5$\,K, see column 9 of Table \ref{tab:results}) at 5\,GHz and 8.4\,GHz would also imply non-thermal radio emission, because a star-forming region can rarely exceed $10^4$\,K at such high frequencies \citep{1992ARA&A..30..575C}. The highest brightness temperatures from star-forming regions have been observed in luminous and ultra-luminous infrared galaxies (LIRGs and ULIRGs, respectively), which is $10^4-10^5$\,K \citep[e.g.][]{1991ApJ...378...65C, 2009A&A...507L..17P, 2014A&A...566A..15V}. As none of these eight sources has been found to be a LIRGs/ULIRGs, thus strongly indicats the AGN dominated radio emission \citep[e.g.][]{1997ApJ...488..621U}. Indeed, among these 8 objects, high radio brightness temperature nuclei ($T_B>10^6$\,K) have been detected for Mrk\,335 and IZw1 by our Very Long Baseline Array (VLBA) observations at 1.4 GHz (Obs. ID: BY145, in preparation) and for Akn\,564 \citep{2004A&A...425...99L}.

We note that more than half of the sources in our sample have a very low radio brightness temperature as compared with an AGN. We have checked the infrared and far-infrared data and found that 8 sources were detected by the Infrared Astronomical Satellite (IRAS). Among them, four sources (B3\,1702+457, Mrk\,957, Mrk\,478 and PG\,1402+261) have the infrared and far-infrared luminosity $L_\mathrm{IR/FIR}>10^{11}L_\odot$, i.e. the so-called type-1 (ultra-) luminous infrared galaxies \citep[e.g.][]{1997ApJ...488..621U}. Among this four objects, the radio emission from B3\,1702+457 has already been proven to be from AGN \citep[][]{2010AJ....139.2612G}, while the radio emission from the remaining sources are likely dominated by the star-forming activity. It should be emphasized here that the radio brightness temperature estimated in this paper is only the lower limit, which is constrained by the limited resolution of VLA A-array. Therefore, high-resolution very long baseline interferometry (VLBI) observation is crucial to distinguish the various origins of the radio emission from radio-quiet AGNs. 

\begin{table*}
\renewcommand\arraystretch{0.8}                                                                                                                                                                              
\centering                                                                                                                                                                                 
\caption{The bolometric luminosities of PG Quasars}  \label{tab:pg}                                                                                                                                                                                                                                                                                                                                                  
\begin{tabular}{lclclclc}                                                                                                                                                         
\toprule

\tc{Name}  &  log\,$L_\mathrm{bol}$  &     \tc{Name}  &  log\,$L_\mathrm{bol}$  &       \tc{Name}  &  log\,$L_\mathrm{bol}$  &     \tc{Name}  &  log\,$L_\mathrm{bol}$        \\
      &  ($\mathrm{erg\,s^{-1}}$)   &           &  ($\mathrm{erg\,s^{-1}}$)   &       &  ($\mathrm{erg\,s^{-1}}$)   &        &  ($\mathrm{erg\,s^{-1}}$)       \\

\tc{(1)}            &     (2)          &     \tc{(3)}            &    (4)          &       \tc{(5)}            &    (6)          &     \tc{(7)}              &     (8)                       \\
\hline
PG\,0007+106   &$    45.38{}^b   $&     PG\,1103-006   &$   46.30{}^a   $&       PG\,1307+085   &$   45.88{}^c   $&     PG\,1534+580     &$    44.43{}^a                $\\
PG\,0026+1298  &$    46.16{}^c   $&     PG\,1114+445   &$   45.39{}^a   $&       PG\,1309+355   &$   45.74{}^a   $&     PG\,1545+210     &$    45.93{}^a                $\\
PG\,0052+251   &$    45.94{}^a   $&     PG\,1116+215   &$   46.31{}^a   $&       PG\,1351+640   &$   45.30{}^a   $&     PG\,1617+175     &$    44.56 - 45.55{}^d        $\\
PG\,0804+761   &$    45.47{}^b   $&     PG\,1202+28    &$   45.41{}^a   $&       PG\,1352+183   &$   45.61{}^a   $&     PG\,2130+099     &$    44.77{}^b                $\\
PG\,0844+349   &$    45.31{}^a   $&     PG\,1211+143   &$   45.77{}^b   $&       PG\,1411+442   &$   45.33{}^a   $&     PG\,2251+113     &$    46.35{}^a                $\\
PG\,0923+129   &$    45.32{}^b   $&     PG\,1216+069   &$   46.31{}^a   $&       PG\,1415+451   &$   45.37{}^a   $&                      &$                             $\\
PG\,0953+414   &$    46.42{}^a   $&     PG\,1229+204   &$   45.65{}^b   $&       PG\,1426+015   &$   45.65{}^b   $&                      &$                             $\\
PG\,1100+772   &$    46.46{}^a   $&     PG\,1259+593   &$   46.83{}^a   $&       PG\,1444+407   &$   46.19{}^a   $&                      &$                             $\\

\hline                                                                                                                                                                            
\end{tabular}

\begin{flushleft}                                                                                                                                                                 
\textbf{References}: (a) \citet{2012MNRAS.426.2677R}; (b) \citet{2017MNRAS.468....2M}; (c) \citet{2007MNRAS.381.1235V}; 
(d) \citet{2016MNRAS.458.1839C}; (e) \citet{2012MNRAS.420.1825J}                                                                                                                                                                                                                                                                                                                                                                                                                                                                                                                                                                                                                                                                                                                                                                                                                                              
\end{flushleft}

\end{table*}                                                                                                                                                                      

\begin{figure*}
\centering
\includegraphics[width=18cm,height=6cm]{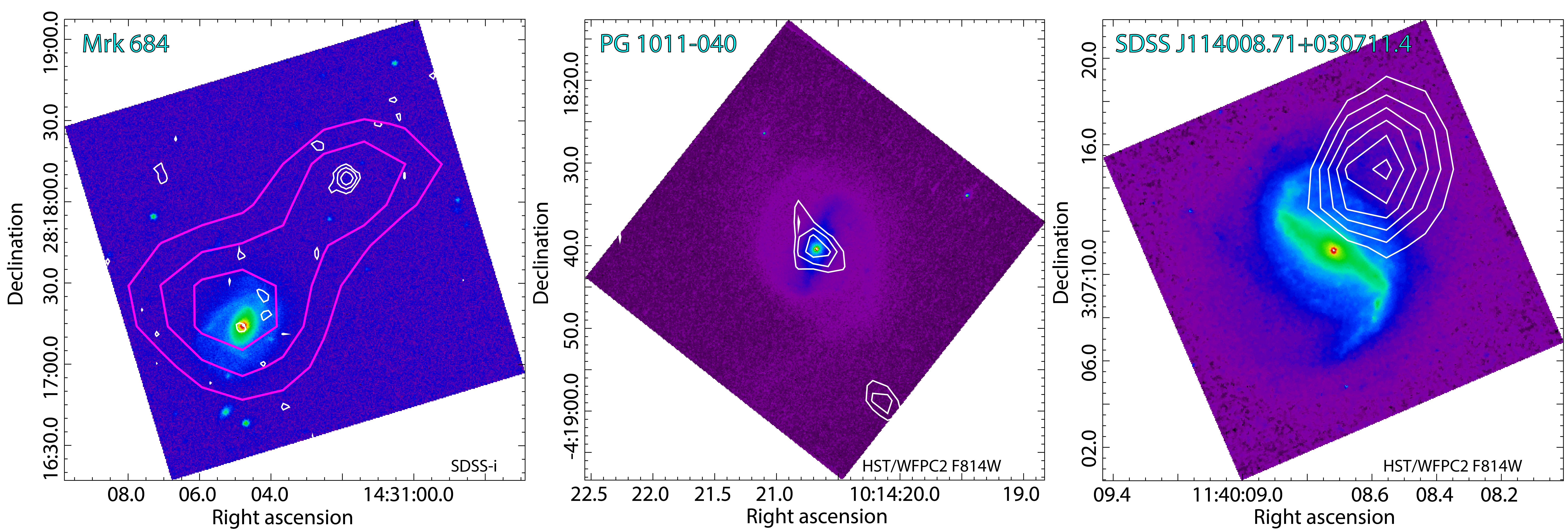}
\caption{The overlaid images of the FIRST (white contours) and/or NVSS (pink contours) 1.4\,GHz VLA images on the optical images (pseudo-color images). The markers for the optical images are at right bottom. The optical image for Mrk\,684 is from the SDSS $i$-band and for PG\,1011-040 and SDSS\,J114008.71+030711.4 are from HST/WFPC F814W. The contours are plotted as (1, 2.25, 3.37, 5.06, ...)$\times3\sigma$.}
\label{fig:ksr}
\end{figure*}

\subsubsection{Estimating star formation rate surface density with radio emission}

In order to further distinguish the radio emission from the star-forming activities and AGN, we can compare the star formation rates (SFRs) measured by using direct indicators with the radio emission estimates. Here we estimate the SFR requirements for the observed radio emission by using the empirical relation derived by \citet{1992ARA&A..30..575C}. It was proven in star-forming galaxies that the SFRs estimated with this relation are well consistent with those estimated from the infrared radiation \citep[e.g., see][]{2014ApJ...780...19R}. Here we only estimate the SFRs from the non-thermal process since there are higher SFRs requirements for thermal processes than for non-thermal to produce equal radio luminosity. This will give us the lowest SFR requirement in producing the observed radio emission. The formula is: 
\begin{equation}
(\frac{L_N}{\mathrm{W\,Hz^{-1}}}) \sim 5.3\times10^{21}(\frac{\nu}{\mathrm{GHz}})^{-0.8}(\frac{\mathrm{SFR}_N(M\geqslant5\,M_\odot)}{M_\odot\,\mathrm{yr^{-1}}}),
\end{equation}
where $L_N$ is the non-thermal radio luminosities, $\mathrm{SFR}_N(M\geqslant5\,M_\odot)$ is the star formation rate of stars more massive than $5\,M_\odot$, and estimated from the non-thermal process (i.e. non-thermal radio flux density). The extended Miller-Scalo initial mass function \citep[][]{1979ApJS...41..513M} with an exponent of $-2.5$ was used, and it was assumed that all stars with mass greater than $8\,M_\odot$ become supernovae, and that the dust absorption is negligible in deriving these formulae \citep{1992ARA&A..30..575C}. Here we assume a typical spectral index of $\alpha_N=-0.8$ for the non-thermal radio emission. The final $\mathrm{SFR}_N(M\geqslant5\,M_\odot)$ is subsequently scaled to the total SFRs ($M\geqslant0.1\,M_\odot$) by a scaling factor of 5.6 \citep[][]{2014ApJ...780...19R}. We calculated the disk-averaged surface densities of the star formation rates ($\Sigma_{SFR}$) by dividing the size of emission regions, derived from the major and minor axis of the synthesis beam. 

The results are listed in the last column of Table \ref{tab:results}. Most of our EESBHs (20 out of 25) have a lower limit for the SFR surface density $\Sigma_\mathrm{SFR}\gtrsim10\,\mathrm{M_\odot\,yr^{-1}\,kpc^{-2}}$ estimated at 5\,GHz or 8.4\,GHz. Among 20 sources, ten have $\Sigma_\mathrm{SFR}\gtrsim100\,\mathrm{M_\odot\,yr^{-1}\,kpc^{-2}}$. It was noted that NLS1s tend to have a high fraction of circumnuclear star-forming rings at more or less 1\,kpc from the core \citep{2006AJ....132..321D, 2018MNRAS.477.1086H}, while it is still very difficult to reach a star formation surface density of $10\,\mathrm{M_\odot\,yr^{-1}\,kpc^{-2}}$ \citep[see][]{2004ARA&A..42..603K, 2012ARA&A..50..531K}. Five objects in our sample have the lowest requirement of $\Sigma_\mathrm{SFR}<10\,\mathrm{M_\odot\,yr^{-1}\,kpc^{-2}}$ in producing the observed radio emission, which are Mrk\,42, Mrk\,684, PG\,1115+407, SDSS\,J010712.04+140845.0 and SDSS\,J114008.71+030711.4. Mrk\,42 and Mrk\,684 have a far-infrared luminosity $L_\mathrm{FIR}\sim10^{10}\,L_\odot$, suggesting a high SFRs on the scale of the whole galaxy, that does not result in a high star formation rate surface density. Particularly, Mrk\,42 was found to have a starburst ring at $\sim300\,\mathrm{pc}$ \citep{2007AJ....134..648M} and have a total ring SFR of $1.38\,\mathrm{M_\odot\,yr^{-1}}$ \citep{2018MNRAS.477.1086H}, which is still below the lower limit required to produce the observed radio emission. Moreover, the VLA A-array has a resolution of $\sim200\,\mathrm{pc}$ at C-band for Mrk\,42, which suggests that the radio emission from Mrk\,42 is less likely to result from the star-forming activity. We should note here that the SFR surface density required to produce the observed radio emission will be even larger than the value given here if we take thermal fractions into account and use the intrinsic size of the emission regions. We thus conclude that the radio emission from most of our EESBHs (at least 21 out of 25) is from AGN activities, otherwise strong nuclear star formation is required to give the observed radio emission.

\subsection{AGNs with super-Eddington accretion rates: comparing with XRBs}

With the population of EESBHs, as presented in this paper, we found that super-Eddington accreting AGNs have a low radio luminosity ($L_5\sim10^{38}\,\mathrm{erg\,s^{-1}}$) and most of them ($\gtrsim90\%$) are radio quiet ($\mathcal{R}\lesssim10$). There is only one super-Eddington source (KUG\,1031+398) in our sample marginally located in the radio-loud region ($\mathcal{R}=12$). This confirms the radio quiescence when Eddington ratios approach or exceed one. More generally, there is an inverse correlation between radio loudness and the Eddington ratio from the sub-Eddington to the super-Eddington regime. In this work, we have extended the inverse $\mathcal{R}-\lambda_{Edd}$ correlation to the super-Eddington regime (see Figure \ref{fig:re}), supporting a continuous jet suppression with an increasing Eddington ratio as proposed by \citet{2002ApJ...564..120H} when overcoming the Eddington limit.

AGNs with extremely high Eddington ratios and Galactic accreting black holes in a very high state \citep[e.g., ][]{2006ApJ...636...56G}, hold the same tendency of the radio loudness versus the Eddington ratio \citep[e.g., ][]{2011MNRAS.417..184B}. Furthermore, with the state transition in XRBs from the low/hard state to the high/soft state, a radio outburst is always associated with the soft X-ray peak at the end of the transition phase, \textit{i.e.}, the very high state. Some of our extremely high Eddington accreting AGNs indeed show kpc-scale emission or kpc-scale radio structures (KSRs) in their FIRST and NVSS images, implying a past ejecting activity. The kpc-scale radio emission in B3\,1702+457 was already noted by \citet{2017A&A...600A..87G}. Mrk\,684 shows an extension of $\sim60$\,kpc to the north-west in the NVSS 1.4\,GHz image (left panel of Figure \ref{fig:ksr}). The NVSS image reveals a two-component system where the second component is consistent in position with FIRST. A visual inspection of the SDSS-$i$ image confirms that there is no obvious optical counterpart associated with the second component that also shows a structure of an AGN hotspot, implying a physical connection with the radio nucleus. PG\,1011-040 shows a slight bipolar extension in the FIRST image towards the north-east and the south-west, perpendicular to the major axis of the host galaxy (middle panel of Figure \ref{fig:ksr}). These characteristics more likely imply that the central AGN is responsible for this radio emission \citep[e.g.][]{2018arXiv181006067Y}, whereas a strong starburst nucleus is also possible \citep[e.g.][]{1991ApJ...369..320S}. In SDSS\,114008.71+030711.4, we find a strong radio source offset north-west from the galaxy center in the FIRST 1.4\,GHz image at a distance of 4.5\,arcsec (7\,kpc) with a peak flux density of $0.84\pm0.15$\,mJy (see Figure \ref{fig:ksr}). There is no optical counterpart coinciding with this radio source in the HST/WFPC2 F814W image, implying that the central AGN or an off nuclear massive black hole \citep[see][]{2020ApJ...888...36R} is responsible for this emission. Previous studies showed that kpc-scale radio structures might be a common feature in radio-loud Seyfert galaxies. There are about 10 NLS1s so far that show kpc-scale radio structures, including one radio-quiet source \citep{2010ApJ...717.1243G, 2012ApJ...760...41D, 2015ApJ...798L..30D, 2015ApJ...800L...8R}. If AGNs have the similar state transition with XRBs, then the kpc-scale radio structures found in our sample might imply the past ejecting activities during the transition state. Similarly, the Galactic X-ray binary SS\,433 has super-Eddington ratio accretion at all times during the outbursts, and the episodes of jet activity are responsible for the formation and shaping of the surrounding elongated radio structures \citep[e.g.][]{2018MNRAS.475.5360B}.

The very high state in Galactic black holes possesses the highest Eddington ratio but is not super-Eddington. It was suggested that super-Eddington accretion in ULXs and some XRBs may indicate a new `ultra-luminous' accretion state \citep[][]{2009MNRAS.397.1836G, 2013MNRAS.435.1758S, 2015MNRAS.448.3374B, 2016AN....337..534R}, which is supported by the finding of a few micro-quasars with super-Eddington accretion that can transit between classical states and the ultra-luminous state \citep[e.g., GRS\,1915+105, see][]{2004MNRAS.355.1105F}. The same scenario was also proposed to explain the powerful radio-emitting outflows in tidal disruption events \citep[e.g., ][]{2011MNRAS.416.2102G}. We propose that a few (if not all) of the super-Eddington accreting AGNs can be a better analogy of stellar-mass accreting systems in the ultra-luminous state, but with longer timescales.

\begin{figure}
\centering
\includegraphics[scale=0.445]{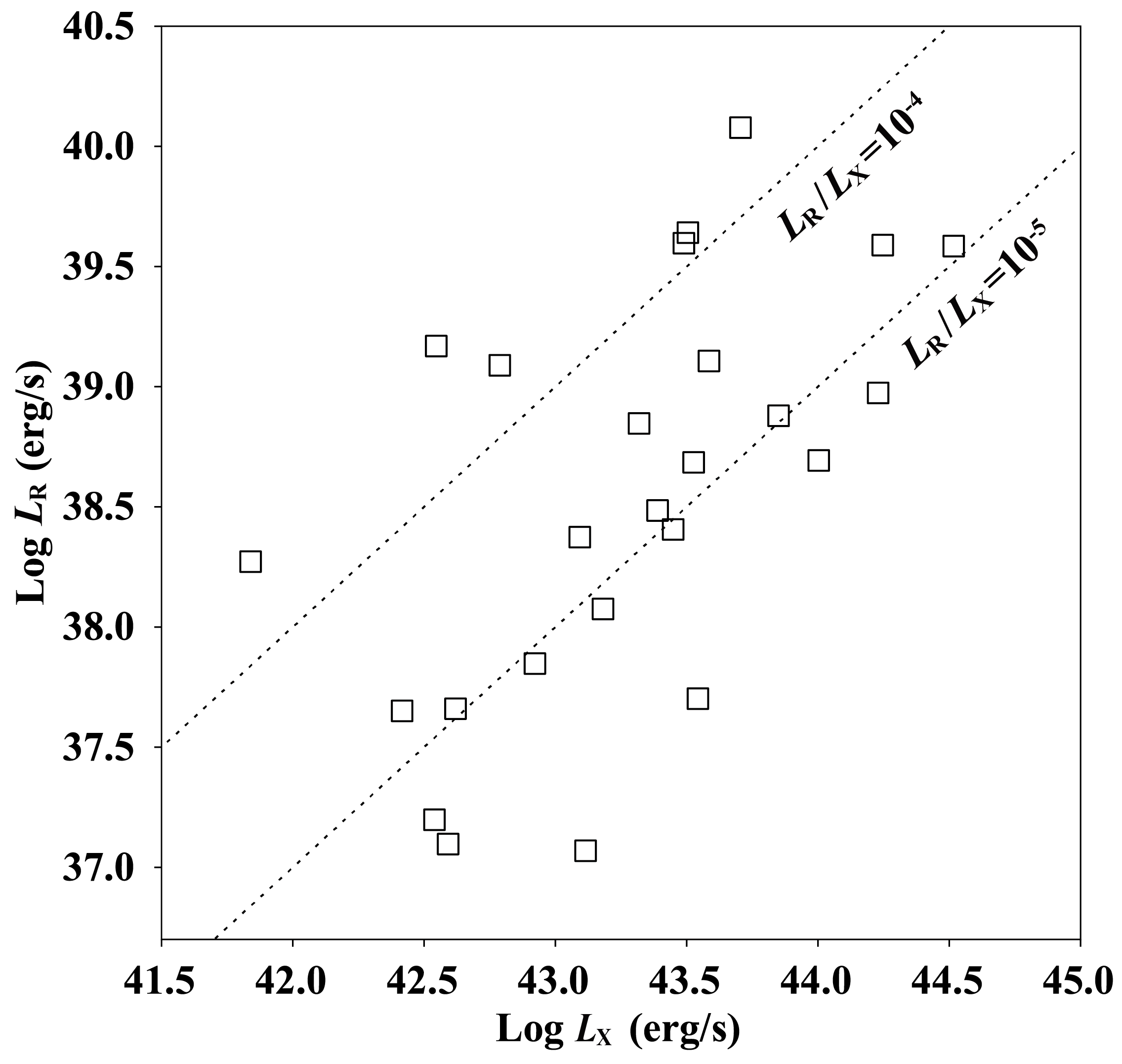}
\caption{The 5\,GHz radio luminosity versus the 2-10\,keV X-ray luminosity for our EESBH sources. The grey dashed lines correspond to $L_\mathrm{R}/L_\mathrm{X}=10^{-5}$ and $L_\mathrm{R}/L_\mathrm{X}=10^{-4}$.}
\label{fig:lrlx}
\end{figure}

\subsection{The Origin of Radio Emission: Jet-driven versus Outflow-driven and Corona-driven}

The extremely high Eddington ratio accreting AGNs are predominately radio-quiet, whereas the radio emission is not completely absent or is dramatically reduced in comparison with the sub-Eddington ratio AGNs (see Figure \ref{fig:re}). In radio-quiet AGNs, several radio-emitting mechanisms are still in competition with each other, such as low-power jets, radio-emitting wind-like outflows, and magnetized corona/jet base. In this paragraph, we will further discuss the origin of radio emission among several AGN dominated activities.

We have explored the correlation between the radio and X-ray luminosities, and between the radio spectral index and the Eddington ratio. Figure \ref{fig:lrlx} shows the radio luminosity versus the X-ray luminosity with markings of the log\,$L_\mathrm{R}/L_\mathrm{X}=-5$ and log\,$L_\mathrm{R}/L_\mathrm{X}=-4$ lines. Most of our extremely high-Eddington accreting AGNs have a radio to X-ray luminosity ratio in the range log\,$L_\mathrm{R}/L_\mathrm{X}=-5$ to $-4$. Sources with a radio to X-ray luminosity ratio log\,$L_\mathrm{R}/L_\mathrm{X}\sim-5$ are found to be similar to coronally active stars, which would imply magnetized corona/jet base dominated radio emission. Because the radio emission from radio-loud AGNs, and hence from jets, have a radio to X-ray luminosity ratio of log\,$L_\mathrm{R}/L_\mathrm{X}\sim10^{-2}$ to $1$, therefore, sources having a radio to X-ray luminosity ratio close to or slightly exceeding log\,$L_\mathrm{R}/L_\mathrm{X}\sim-4$ would imply a combination of corona and jets \citep[][]{2008MNRAS.390..847L, 2015MNRAS.451..517B, 2019MNRAS.482.5513L}.

\begin{figure}
\centering
\includegraphics[scale=0.32]{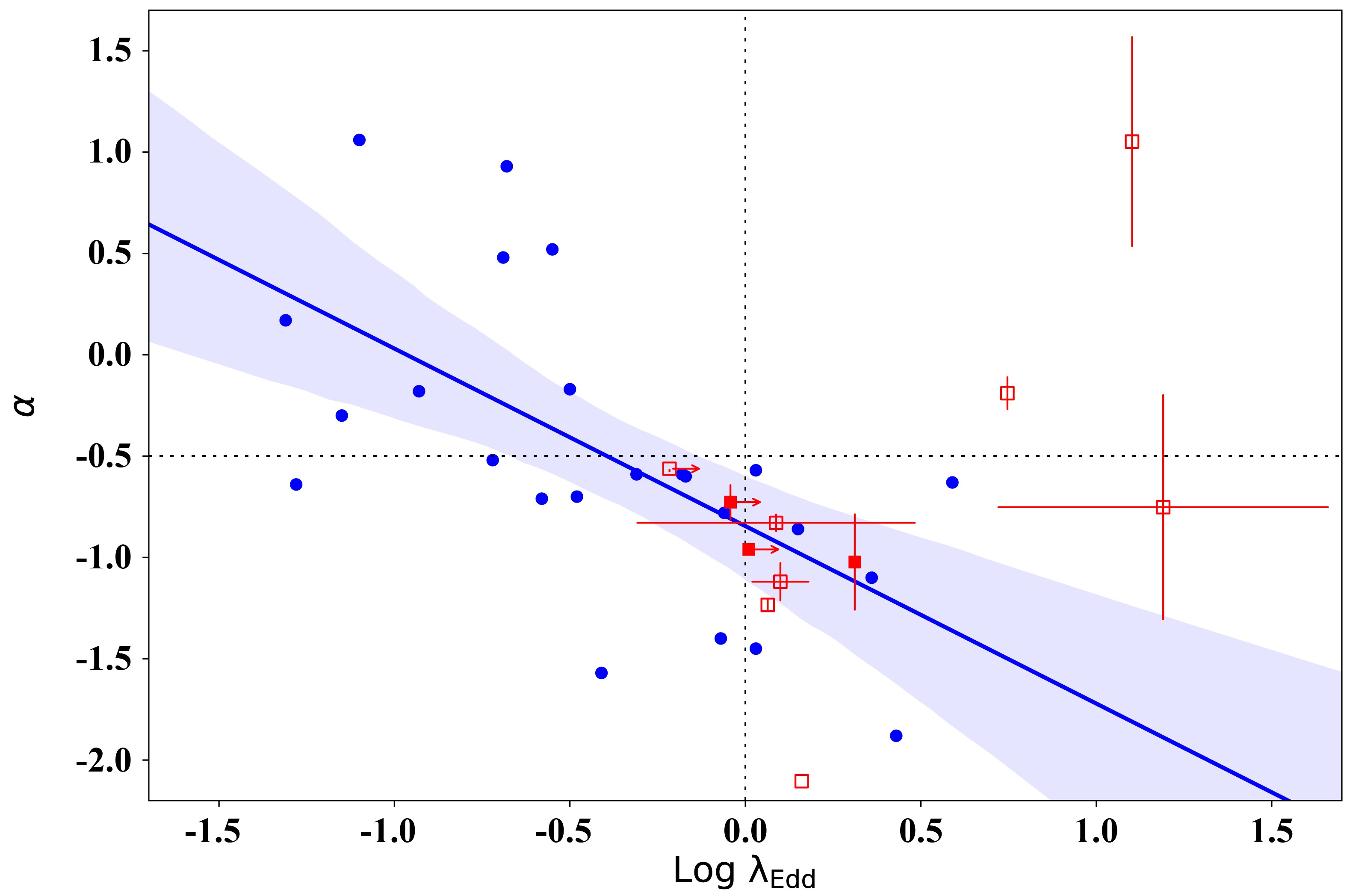}
\caption{The radio spectral index $\alpha$ versus Eddington ratio $\lambda_\mathrm{Edd}$. The spectral index is measured between 1.4 and 5\,GHz (red filled squares) and between 5 and 8.5\,GHz (red unfilled squares) for our EESBHs. 5 $\sim$ 8.4\,GHz spectral index and the Eddington ratio distribution for radio-quiet quasars from \citet{2019MNRAS.482.5513L} are plotted as blue dots, where the blue line and belt are linear regression and 95\% confidence interval, respectively. The horizontal dotted line at $\alpha=-0.5$ is the division of flat (above) and steep (below) spectrum radio sources, the vertical dotted line designates $\lambda_\mathrm{Edd}=1$.}
\label{fig:siedd}
\end{figure}

Figure \ref{fig:siedd} shows the distribution of the non-simultaneous radio spectral index $\alpha$ along the Eddington ratio $\lambda_\mathrm{Edd}$, the spectral index uncertainties are estimated from the integrated flux density errors. The non-simultaneity of the 1.4, 5 and 8.4\,GHz observations in these objects may induce spectral index errors, whereas it may not lead to a large uncertainty in spectral index due to the stability of radio emission of our EESBHs (see Section \ref{sec:results1}). Most of our (9 out of 11) sources have a steep radio spectrum, implying that the radio emission from our EESBHs is dominated by the ejecta rather than the core. In Figure \ref{fig:siedd}, we also plot the $5 - 8.4$\,GHz spectral index and the Eddington ratio distribution for radio-quiet quasars from \citet[][]{2019MNRAS.482.5513L} as a comparison. \citet{2019MNRAS.482.5513L} discovered an inverse correlation between $\alpha$ and $\lambda_\mathrm{Edd}$ using a sample of 25 radio-quiet quasars, spanning a large range of log$\lambda_\mathrm{Edd}$ from $-$1.6 to 0.6. Note that the high-Eddington (log\,$\lambda_\mathrm{Edd}>-0.5$) quasars tend to have a steep spectrum ($\alpha<-0.5$, \textit{i.e.}, below the horizontal dotted line in Figure \ref{fig:siedd}), and it was interpreted as that the radio emission is from optically thin outflows \citep[see][]{2019MNRAS.482.5513L}. It is clear from Figure \ref{fig:siedd} that most of our high-Eddington sources follow a similar trend. Furthermore, the 11 EESBHs having a radio to X-ray luminosity ratio of log\,$L_\mathrm{R}/L_\mathrm{X}\sim-5$, suggesting the contribution from the radio-emitting corona \citep[][]{2008MNRAS.390..847L, 2015MNRAS.451..517B, 2019MNRAS.482.5513L}. To sum up, a possible interpretation for the radio emission in our EESBHs is the transient ejecta caused by the optically thin outflow (outflowing corona), which enhance the comparison between the high Eddington AGNs and XRBs in the soft state since the transient ejections were generally observed in the soft state of XRBs.

Interestingly, few of our highly super-Eddington systems with log\,$\lambda_\mathrm{Edd} \gtrsim 0.6$ do not follow the above trend (see Figure \ref{fig:siedd}). The flatter radio spectra in such cases might be likely due to the jet activity. However, the observed radio to X-ray luminosity ratio for the three highly super-Eddington ratio AGNs is $L_\mathrm{R}/L_\mathrm{X}\lesssim10^{-5}$, which rules out the possibility of traditional (`BP' or `BZ') jets \citep[e.g.][]{1977MNRAS.179..433B, 1982MNRAS.199..883B} and the radio emission is suggested to be dominated by a magnetized corona (aka jet base). The other possibility is the radiation-pressure-driven jets start to dominate in this regime \citep[e.g.][]{2009PASJ...61..783T, 2015MNRAS.453.3213S, 2015NatPh..11..551F}, where a weak and short jet produces optically thick radio emission and appears like a scaled-down version of radio-loud AGNs. Interestingly, there is a slight positive $\alpha$ - $\lambda_\mathrm{Edd}$ correlation has been found in radio-loud quasars \citep[see][]{2019MNRAS.482.5513L}, which may likely support this idea. Extended radio emission was observed in the Galactic microquasar SS\,433, which was attributed to a mixture of radiation-pressure-driven jets and radio-emitting outflow \citep[e.g.][]{2005MNRAS.357..295O}. There is only one source (Akn\,564) in our sample that shows such quasi-continuous radio ejection, and VLBA images show that B3\,1702+457 in our EESBHs sample has a bipolar emission \citep{2011ApJ...738..126D}. More such super-Eddington accreting AGNs with simultaneous multi-band observations are required to confirm these spectral characteristics. Furthermore, future VLBI observation of super-Eddington accreting AGNs will provide more evidence on the different radio-emitting mechanisms as proposed here.

\section{Summary}
The main results of this paper are summarized as follow:

1. The compact and core-dominated radio emission is found in our sample of EESBHs. Most of the radio emission originates within the central few hundred parsecs to one kilo-parsec region, implying that the sources are compact at these scales. 

2. In our extremely high Eddington accreting systems, we have estimated the lowest star formation rate surface density required to produce the observed radio emission. This surface density is higher than the maximum value that has been detected in circumnuclear starburst galaxies suggesting that the radio emission comes from AGN activity. 

3. A global inverse correlation has been established between the radio loudness $\mathcal{R}$ and the Eddington ratio $\lambda_\mathrm{Edd}$ from the sub- to the super-Eddington regime. There is no clear demarcation found in this distribution for EESBHs, indicating a continuous suppression of radio emission from the high- to the super-Eddington regime. Super-Eddington accreting AGNs are predominately radio-quiet, but not completely radio quenched. 

4. The high-Eddington and mildly super-Eddington AGNs ($-0.5<$log\,$\lambda_\mathrm{Edd}<0.6$) have a radio to X-ray luminosity ratio $L_\mathrm{R}/L_\mathrm{X}\sim10^{-5} $ -- $ 10^{-4}$ and a steep radio spectrum, implying the radio emission was dominated by the transient ejecta where the outflowing corona may be at work.

\acknowledgments

This work was supported by the National Key R\&D Program of China (2016YFA0400702) and the  National Science Foundation of China (11721303, 11991052). SY thanks for the support from the KIAA-CAS fellowship, which is jointly supported by Peking
University and Chinese Academy of Sciences. SY is supported by the Boya fellowship. 
MFG acknowledges support from the National Science Foundation of China (11873073). 
This work makes use of public data from NSF's Karl G. Jansky Very Large Array (VLA), the VLA facility is operated by 
National Radio Astronomy Observatory (NRAO). The National Radio Astronomy Observatory 
is a facility of the National Science Foundation operated under cooperative agreement
by Associated Universities, Inc. Some of the data presented in this paper were obtained from 
the Mikulski Archive for Space Telescopes (MAST). STScI is operated by the Association of 
Universities for Research in Astronomy, Inc., under NASA contract NAS5-26555. This work 
makes use of SDSS data, Funding for the Sloan Digital Sky Survey IV has been provided by 
the Alfred P. Sloan Foundation, the U.S. Department of Energy Office of Science, and the 
Participating Institutions. SDSS acknowledges support and resources from the Center for 
High-Performance Computing at the University of Utah. 

%

\appendix
\section{NOTES ON VLA RESULTS FOR A FEW SOURCES} \label{appendix}
\textbf{Akn\,564}: We have measured a peak flux density of $5.69 \pm 0.03$\,mJy\,beam$^{-1}$ at C band, which is consistent with the peak 
flux density of $5.68\pm0.02$\,mJy\,beam$^{-1}$ obtained by \citet{2018A&A...614A..87B}. The source is resolved into three components at C and X bands, we only take account of the integrated flux density of the central component, which is $5.84\pm0.03$\,mJy at C band and $3.18\pm0.07$\,mJy at X band. \citet{2001ApJS..132..199S} also fit the X-band image with three components, and the inferred integrated flux density of the central component is 3.1\,mJy, and being consistent with our results. The size of the radio structure is $\sim2$\,arcsec (corresponding to physical linear size of $\sim 4$\,kpc). 

\textbf{Mrk\,684}: The VLA A-array observation at C band has the baseline range from $0$ to 0.6\,M$\lambda$. Here we constrain the baseline to be $>0.04$\,M$\lambda$, this procedure rejects some diffused emission and results in detection of the central source with a peak flux density of $\sim0.05$\,mJy beam$^{-1}$, but the signal-to-noise ratio is only 3.3. 

\textbf{IRASF\,12397+3333}: This source shows diffused emission with a linear extension along the major axis of its host galaxy. This radio structure can be modeled with four Gaussian components. Here we only take into account the flux density from the central compact component as the integrated flux of this source. 

\textbf{IRAS\,04416+1215}: There is no available VLA data at C band for this source. Alternatively, we estimated a C-band flux density by using the L-band to X-band spectral index ($\alpha_{1.4}^{8.4}=-0.91$), the resulted 5\,GHz flux density is 4.07\,mJy. 

\textbf{SDSS\,J010712.04+140845.0} and \textbf{PG\,1115+407}: These two sources are not  detected at C band. We take the three times \textit{rms} noise as the upper limit of the flux density. 

\textbf{SDSS\,J114008.71+030711.4}: This source is marginally detected with an SNR of $\sim3\sigma$. 




\bibliographystyle{aasjournal}
\bibliography{eesbhs}

\begin{thebibliography}{}
\expandafter\ifx\csname natexlab\endcsname\relax\def\natexlab#1{#1}\fi
\providecommand{\url}[1]{\href{#1}{#1}}
\providecommand{\dodoi}[1]{doi:~\href{http://doi.org/#1}{\nolinkurl{#1}}}
\providecommand{\doeprint}[1]{\href{http://ascl.net/#1}{\nolinkurl{http://ascl.net/#1}}}
\providecommand{\doarXiv}[1]{\href{https://arxiv.org/abs/#1}{\nolinkurl{https://arxiv.org/abs/#1}}}

\bibitem[{{Abdo} {et~al.}(2009){Abdo}, {Ackermann}, {Ajello}, {Baldini},
  {Ballet}, {Barbiellini}, {Bastieri}, {Bechtol}, {Bellazzini}, {Berenji}, \&
  et~al.}]{2009ApJ...707L.142A}
{Abdo}, A.~A., {Ackermann}, M., {Ajello}, M., {et~al.} 2009, ApJ, 707, L142

\bibitem[{{Abramowicz} {et~al.}(1988){Abramowicz}, {Czerny}, {Lasota}, \&
  {Szuszkiewicz}}]{1988ApJ...332..646A}
{Abramowicz}, M.~A., {Czerny}, B., {Lasota}, J.~P., \& {Szuszkiewicz}, E. 1988,
  \apj, 332, 646

\bibitem[{{Alberdi} {et~al.}(2006){Alberdi}, {Colina}, {Torrelles}, {Panagia},
  {Wilson}, \& {Garrington}}]{2006ApJ...638..938A}
{Alberdi}, A., {Colina}, L., {Torrelles}, J.~M., {et~al.} 2006, \apj, 638, 938

\bibitem[{{Auchettl} {et~al.}(2017){Auchettl}, {Guillochon}, \&
  {Ramirez-Ruiz}}]{2017ApJ...838..149A}
{Auchettl}, K., {Guillochon}, J., \& {Ramirez-Ruiz}, E. 2017, \apj, 838, 149

\bibitem[{{Batejat} {et~al.}(2011){Batejat}, {Conway}, {Hurley}, {Parra},
  {Diamond}, {Lonsdale}, \& {Lonsdale}}]{2011ApJ...740...95B}
{Batejat}, F., {Conway}, J.~E., {Hurley}, R., {et~al.} 2011, \apj, 740, 95

\bibitem[{{Becker} {et~al.}(1995){Becker}, {White}, \&
  {Helfand}}]{1995ApJ...450..559B}
{Becker}, R.~H., {White}, R.~L., \& {Helfand}, D.~J. 1995, \apj, 450, 559

\bibitem[{{Begelman} {et~al.}(2006){Begelman}, {King}, \&
  {Pringle}}]{2006MNRAS.370..399B}
{Begelman}, M.~C., {King}, A.~R., \& {Pringle}, J.~E. 2006, \mnras, 370, 399

\bibitem[{{Begelman} \& {Volonteri}(2017)}]{2017MNRAS.464.1102B}
{Begelman}, M.~C., \& {Volonteri}, M. 2017, \mnras, 464, 1102

\bibitem[{{Behar} {et~al.}(2015){Behar}, {Baldi}, {Laor}, {Horesh}, {Stevens},
  \& {Tzioumis}}]{2015MNRAS.451..517B}
{Behar}, E., {Baldi}, R.~D., {Laor}, A., {et~al.} 2015, \mnras, 451, 517

\bibitem[{{Behar} {et~al.}(2018){Behar}, {Vogel}, {Baldi}, {Smith}, \&
  {Mushotzky}}]{2018MNRAS.478..399B}
{Behar}, E., {Vogel}, S., {Baldi}, R.~D., {Smith}, K.~L., \& {Mushotzky}, R.~F.
  2018, \mnras, 478, 399

\bibitem[{{Berger} {et~al.}(2012){Berger}, {Zauderer}, {Pooley}, {Soderberg},
  {Sari}, {Brunthaler}, \& {Bietenholz}}]{2012ApJ...748...36B}
{Berger}, E., {Zauderer}, A., {Pooley}, G.~G., {et~al.} 2012, \apj, 748, 36

\bibitem[{{Berton} {et~al.}(2018){Berton}, {Congiu}, {J{\"a}rvel{\"a}},
  {Antonucci}, {Kharb}, {Lister}, {Tarchi}, {Caccianiga}, {Chen}, {Foschini},
  {L{\"a}hteenm{\"a}ki}, {Richards}, {Ciroi}, {Cracco}, {Frezzato}, {La Mura},
  \& {Rafanelli}}]{2018A&A...614A..87B}
{Berton}, M., {Congiu}, E., {J{\"a}rvel{\"a}}, E., {et~al.} 2018, \aap, 614,
  A87

\bibitem[{{Blandford} \& {Payne}(1982)}]{1982MNRAS.199..883B}
{Blandford}, R.~D., \& {Payne}, D.~G. 1982, \mnras, 199, 883

\bibitem[{{Blandford} \& {Znajek}(1977)}]{1977MNRAS.179..433B}
{Blandford}, R.~D., \& {Znajek}, R.~L. 1977, \mnras, 179, 433

\bibitem[{{Bloom} {et~al.}(2011){Bloom}, {Giannios}, {Metzger}, {Cenko},
  {Perley}, {Butler}, {Tanvir}, {Levan}, {O'Brien}, {Strubbe}, {De Colle},
  {Ramirez-Ruiz}, {Lee}, {Nayakshin}, {Quataert}, {King}, {Cucchiara},
  {Guillochon}, {Bower}, {Fruchter}, {Morgan}, \& {van der
  Horst}}]{2011Sci...333..203B}
{Bloom}, J.~S., {Giannios}, D., {Metzger}, B.~D., {et~al.} 2011, Science, 333,
  203

\bibitem[{{Blundell} {et~al.}(2001){Blundell}, {Mioduszewski}, {Muxlow},
  {Podsiadlowski}, \& {Rupen}}]{2001ApJ...562L..79B}
{Blundell}, K.~M., {Mioduszewski}, A.~J., {Muxlow}, T. W.~B., {Podsiadlowski},
  P., \& {Rupen}, M.~P. 2001, \apj, 562, L79

\bibitem[{{Bonzini} {et~al.}(2013){Bonzini}, {Padovani}, {Mainieri},
  {Kellermann}, {Miller}, {Rosati}, {Tozzi}, \&
  {Vattakunnel}}]{2013MNRAS.436.3759B}
{Bonzini}, M., {Padovani}, P., {Mainieri}, V., {et~al.} 2013, \mnras, 436, 3759

\bibitem[{{Boroson} \& {Green}(1992)}]{BG92}
{Boroson}, T.~A., \& {Green}, R.~F. 1992, ApJS, 80, 109

\bibitem[{{Brightman} {et~al.}(2013){Brightman}, {Silverman}, {Mainieri},
  {Ueda}, {Schramm}, {Matsuoka}, {Nagao}, {Steinhardt}, {Kartaltepe},
  {Sanders}, {Treister}, {Shemmer}, {Brandt}, {Brusa}, {Comastri}, {Ho},
  {Lanzuisi}, {Lusso}, {Nandra}, {Salvato}, {Zamorani}, {Akiyama}, {Alexander},
  {Bongiorno}, {Capak}, {Civano}, {Del Moro}, {Doi}, {Elvis}, {Hasinger},
  {Laird}, {Masters}, {Mignoli}, {Ohta}, {Schawinski}, \&
  {Taniguchi}}]{2013MNRAS.433.2485B}
{Brightman}, M., {Silverman}, J.~D., {Mainieri}, V., {et~al.} 2013, \mnras,
  433, 2485

\bibitem[{{Broderick} \& {Fender}(2011)}]{2011MNRAS.417..184B}
{Broderick}, J.~W., \& {Fender}, R.~P. 2011, \mnras, 417, 184

\bibitem[{{Broderick} {et~al.}(2018){Broderick}, {Fender}, {Miller-Jones},
  {Trushkin}, {Stewart}, {Anderson}, {Staley}, {Blundell}, {Pietka}, {Markoff},
  {Rowlinson}, {Swinbank}, {van der Horst}, {Bell}, {Breton}, {Carbone},
  {Corbel}, {Eisl{\"o}ffel}, {Falcke}, {Grie{\ss}meier}, {Hessels},
  {Kondratiev}, {Law}, {Molenaar}, {Serylak}, {Stappers}, {van Leeuwen},
  {Wijers}, {Wijnands}, {Wise}, \& {Zarka}}]{2018MNRAS.475.5360B}
{Broderick}, J.~W., {Fender}, R.~P., {Miller-Jones}, J.~C.~A., {et~al.} 2018,
  \mnras, 475, 5360

\bibitem[{{Brorby} {et~al.}(2015){Brorby}, {Kaaret}, \&
  {Feng}}]{2015MNRAS.448.3374B}
{Brorby}, M., {Kaaret}, P., \& {Feng}, H. 2015, \mnras, 448, 3374

\bibitem[{{Burrows} {et~al.}(2011){Burrows}, {Kennea}, {Ghisellini}, {Mangano},
  {Zhang}, {Page}, {Eracleous}, {Romano}, {Sakamoto}, {Falcone}, {Osborne},
  {Campana}, {Beardmore}, {Breeveld}, {Chester}, {Corbet}, {Covino},
  {Cummings}, {D'Avanzo}, {D'Elia}, {Esposito}, {Evans}, {Fugazza}, {Gelbord},
  {Hiroi}, {Holland}, {Huang}, {Im}, {Israel}, {Jeon}, {Jeon}, {Jun}, {Kawai},
  {Kim}, {Krimm}, {Marshall}, {P. M{\'e}sz{\'a}ros}, {Negoro}, {Omodei},
  {Park}, {Perkins}, {Sugizaki}, {Sung}, {Tagliaferri}, {Troja}, {Ueda},
  {Urata}, {Usui}, {Antonelli}, {Barthelmy}, {Cusumano}, {Giommi}, {Melandri},
  {Perri}, {Racusin}, {Sbarufatti}, {Siegel}, \&
  {Gehrels}}]{2011Natur.476..421B}
{Burrows}, D.~N., {Kennea}, J.~A., {Ghisellini}, G., {et~al.} 2011, \nat, 476,
  421

\bibitem[{{Castell{\'o}-Mor} {et~al.}(2016){Castell{\'o}-Mor}, {Netzer}, \&
  {Kaspi}}]{2016MNRAS.458.1839C}
{Castell{\'o}-Mor}, N., {Netzer}, H., \& {Kaspi}, S. 2016, \mnras, 458, 1839

\bibitem[{{Colina} {et~al.}(2001){Colina}, {Alberdi}, {Torrelles}, {Panagia},
  \& {Wilson}}]{2001ApJ...553L..19C}
{Colina}, L., {Alberdi}, A., {Torrelles}, J.~M., {Panagia}, N., \& {Wilson},
  A.~S. 2001, \apj, 553, L19

\bibitem[{{Condon}(1992)}]{1992ARA&A..30..575C}
{Condon}, J.~J. 1992, Annual Review of Astronomy and Astrophysics, 30, 575

\bibitem[{{Condon} {et~al.}(1998){Condon}, {Cotton}, {Greisen}, {Yin},
  {Perley}, {Taylor}, \& {Broderick}}]{1998AJ....115.1693C}
{Condon}, J.~J., {Cotton}, W.~D., {Greisen}, E.~W., {et~al.} 1998, \aj, 115,
  1693

\bibitem[{{Condon} {et~al.}(1991){Condon}, {Huang}, {Yin}, \&
  {Thuan}}]{1991ApJ...378...65C}
{Condon}, J.~J., {Huang}, Z.~P., {Yin}, Q.~F., \& {Thuan}, T.~X. 1991, \apj,
  378, 65

\bibitem[{{Condon} {et~al.}(1992){Condon}, {Huang}, {Yin}, \&
  {Thuan}}]{1992ASPC...31...79C}
{Condon}, J.~J., {Huang}, Z.~P., {Yin}, Q.~F., \& {Thuan}, T.~X. 1992, in
  Relationships Between Active Galactic Nuclei and Starburst Galaxies, Vol.~31,
  79

\bibitem[{{Dai} {et~al.}(2018){Dai}, {McKinney}, {Roth}, {Ramirez-Ruiz}, \&
  {Miller}}]{2018ApJ...859L..20D}
{Dai}, L., {McKinney}, J.~C., {Roth}, N., {Ramirez-Ruiz}, E., \& {Miller},
  M.~C. 2018, \apj, 859, L20

\bibitem[{{Deo} {et~al.}(2006){Deo}, {Crenshaw}, \&
  {Kraemer}}]{2006AJ....132..321D}
{Deo}, R.~P., {Crenshaw}, D.~M., \& {Kraemer}, S.~B. 2006, \aj, 132, 321

\bibitem[{{Doi} {et~al.}(2013){Doi}, {Asada}, {Fujisawa}, {Nagai}, {Hagiwara},
  {Wajima}, \& {Inoue}}]{2013ApJ...765...69D}
{Doi}, A., {Asada}, K., {Fujisawa}, K., {et~al.} 2013, \apj, 765, 69

\bibitem[{{Doi} {et~al.}(2011){Doi}, {Asada}, \& {Nagai}}]{2011ApJ...738..126D}
{Doi}, A., {Asada}, K., \& {Nagai}, H. 2011, \apj, 738, 126

\bibitem[{{Doi} {et~al.}(2012){Doi}, {Nagira}, {Kawakatu}, {Kino}, {Nagai}, \&
  {Asada}}]{2012ApJ...760...41D}
{Doi}, A., {Nagira}, H., {Kawakatu}, N., {et~al.} 2012, \apj, 760, 41

\bibitem[{{Doi} {et~al.}(2015){Doi}, {Wajima}, {Hagiwara}, \&
  {Inoue}}]{2015ApJ...798L..30D}
{Doi}, A., {Wajima}, K., {Hagiwara}, Y., \& {Inoue}, M. 2015, \apj, 798, L30

\bibitem[{{Dotan} \& {Shaviv}(2011)}]{2011MNRAS.413.1623D}
{Dotan}, C., \& {Shaviv}, N.~J. 2011, \mnras, 413, 1623

\bibitem[{{Du} {et~al.}(2015){Du}, {Hu}, {Lu}, {Huang}, {Cheng}, {Qiu}, {Li},
  {Zhang}, {Fan}, {Bai}, {Bian}, {Yuan}, {Kaspi}, {Ho}, {Netzer}, {Wang}, \&
  {SEAMBH Collaboration}}]{2015ApJ...806...22D}
{Du}, P., {Hu}, C., {Lu}, K.-X., {et~al.} 2015, \apj, 806, 22

\bibitem[{{Fabrika} {et~al.}(2015){Fabrika}, {Ueda}, {Vinokurov}, {Sholukhova},
  \& {Shidatsu}}]{2015NatPh..11..551F}
{Fabrika}, S., {Ueda}, Y., {Vinokurov}, A., {Sholukhova}, O., \& {Shidatsu}, M.
  2015, Nature Physics, 11, 551

\bibitem[{{Falcke} {et~al.}(2004){Falcke}, {K{\"o}rding}, \&
  {Markoff}}]{2004A&A...414..895F}
{Falcke}, H., {K{\"o}rding}, E., \& {Markoff}, S. 2004, \aap, 414, 895

\bibitem[{{Fender} {et~al.}(1999){Fender}, {Corbel}, {Tzioumis}, {McIntyre},
  {Campbell-Wilson}, {Nowak}, {Sood}, {Hunstead}, {Harmon}, {Durouchoux}, \&
  {Heindl}}]{1999ApJ...519L.165F}
{Fender}, R., {Corbel}, S., {Tzioumis}, T., {et~al.} 1999, \apj, 519, L165

\bibitem[{{Fender} {et~al.}(2004){Fender}, {Belloni}, \&
  {Gallo}}]{2004MNRAS.355.1105F}
{Fender}, R.~P., {Belloni}, T.~M., \& {Gallo}, E. 2004, \mnras, 355, 1105

\bibitem[{{Fenech} {et~al.}(2008){Fenech}, {Muxlow}, {Beswick}, {Pedlar}, \&
  {Argo}}]{2008MNRAS.391.1384F}
{Fenech}, D.~M., {Muxlow}, T.~W.~B., {Beswick}, R.~J., {Pedlar}, A., \& {Argo},
  M.~K. 2008, \mnras, 391, 1384

\bibitem[{{Giannios} \& {Metzger}(2011)}]{2011MNRAS.416.2102G}
{Giannios}, D., \& {Metzger}, B.~D. 2011, \mnras, 416, 2102

\bibitem[{{Gies} {et~al.}(2002){Gies}, {McSwain}, {Riddle}, {Wang}, {Wiita}, \&
  {Wingert}}]{2002ApJ...566.1069G}
{Gies}, D.~R., {McSwain}, M.~V., {Riddle}, R.~L., {et~al.} 2002, \apj, 566,
  1069

\bibitem[{{Giroletti} \& {Panessa}(2009)}]{2009ApJ...706L.260G}
{Giroletti}, M., \& {Panessa}, F. 2009, \apj, 706, L260

\bibitem[{{Giroletti} {et~al.}(2017){Giroletti}, {Panessa}, {Longinotti},
  {Krongold}, {Guainazzi}, {Costantini}, \&
  {Santos-Lleo}}]{2017A&A...600A..87G}
{Giroletti}, M., {Panessa}, F., {Longinotti}, A.~L., {et~al.} 2017, \aap, 600,
  A87

\bibitem[{{Gladstone} {et~al.}(2009){Gladstone}, {Roberts}, \&
  {Done}}]{2009MNRAS.397.1836G}
{Gladstone}, J.~C., {Roberts}, T.~P., \& {Done}, C. 2009, \mnras, 397, 1836

\bibitem[{{Gliozzi} {et~al.}(2010){Gliozzi}, {Papadakis}, {Grupe}, {Brinkmann},
  {Raeth}, \& {Kedziora-Chudczer}}]{2010ApJ...717.1243G}
{Gliozzi}, M., {Papadakis}, I.~E., {Grupe}, D., {et~al.} 2010, \apj, 717, 1243

\bibitem[{{Greene} {et~al.}(2006){Greene}, {Ho}, \&
  {Ulvestad}}]{2006ApJ...636...56G}
{Greene}, J.~E., {Ho}, L.~C., \& {Ulvestad}, J.~S. 2006, \apj, 636, 56

\bibitem[{{Greiner} {et~al.}(2001){Greiner}, {Cuby}, \&
  {McCaughrean}}]{2001Natur.414..522G}
{Greiner}, J., {Cuby}, J.~G., \& {McCaughrean}, M.~J. 2001, \nat, 414, 522

\bibitem[{{Gu} \& {Chen}(2010)}]{2010AJ....139.2612G}
{Gu}, M., \& {Chen}, Y. 2010, \aj, 139, 2612

\bibitem[{{G{\"u}ltekin} {et~al.}(2014){G{\"u}ltekin}, {Cackett}, {King},
  {Miller}, \& {Pinkney}}]{2014ApJ...788L..22G}
{G{\"u}ltekin}, K., {Cackett}, E.~M., {King}, A.~L., {Miller}, J.~M., \&
  {Pinkney}, J. 2014, \apj, 788, L22

\bibitem[{{Haardt} \& {Maraschi}(1991)}]{1991ApJ...380L..51H}
{Haardt}, F., \& {Maraschi}, L. 1991, \apj, 380, L51

\bibitem[{{Hada} {et~al.}(2018){Hada}, {Doi}, {Wajima}, {D'Ammando}, {Orienti},
  {Giroletti}, {Giovannini}, {Nakamura}, \& {Asada}}]{2018arXiv180508299H}
{Hada}, K., {Doi}, A., {Wajima}, K., {et~al.} 2018, ArXiv e-prints

\bibitem[{{Hennig} {et~al.}(2018){Hennig}, {Riffel}, {Dors}, {Riffel},
  {Storchi-Bergmann}, \& {Colina}}]{2018MNRAS.477.1086H}
{Hennig}, M.~G., {Riffel}, R.~A., {Dors}, O.~L., {et~al.} 2018, \mnras, 477,
  1086

\bibitem[{{Ho}(2002)}]{2002ApJ...564..120H}
{Ho}, L.~C. 2002, \apj, 564, 120

\bibitem[{{Ho}(2008)}]{2008ARA&A..46..475H}
---. 2008, Annual Review of Astronomy and Astrophysics, 46, 475

\bibitem[{{Hopkins} {et~al.}(2003){Hopkins}, {Afonso}, {Chan}, {Cram},
  {Georgakakis}, \& {Mobasher}}]{2003AJ....125..465H}
{Hopkins}, A.~M., {Afonso}, J., {Chan}, B., {et~al.} 2003, \aj, 125, 465

\bibitem[{{Hovatta} {et~al.}(2012){Hovatta}, {Lister}, {Aller}, {Aller},
  {Homan}, {Kovalev}, {Pushkarev}, \& {Savolainen}}]{2012AJ....144..105H}
{Hovatta}, T., {Lister}, M.~L., {Aller}, M.~F., {et~al.} 2012, \aj, 144, 105

\bibitem[{{Inoue} \& {Doi}(2018)}]{2018ApJ...869..114I}
{Inoue}, Y., \& {Doi}, A. 2018, \apj, 869, 114

\bibitem[{{Jiang} {et~al.}(2017){Jiang}, {Stone}, \&
  {Davis}}]{2017arXiv170902845J}
{Jiang}, Y.-F., {Stone}, J., \& {Davis}, S.~W. 2017, ArXiv e-prints,
  arXiv:1709.02845

\bibitem[{{Jin} {et~al.}(2017){Jin}, {Done}, {Ward}, \&
  {Gardner}}]{2017MNRAS.471..706J}
{Jin}, C., {Done}, C., {Ward}, M., \& {Gardner}, E. 2017, \mnras, 471, 706

\bibitem[{{Jin} {et~al.}(2012{\natexlab{a}}){Jin}, {Ward}, \&
  {Done}}]{2012MNRAS.425..907J}
{Jin}, C., {Ward}, M., \& {Done}, C. 2012{\natexlab{a}}, \mnras, 425, 907

\bibitem[{{Jin} {et~al.}(2012{\natexlab{b}}){Jin}, {Ward}, {Done}, \&
  {Gelbord}}]{2012MNRAS.420.1825J}
{Jin}, C., {Ward}, M., {Done}, C., \& {Gelbord}, J. 2012{\natexlab{b}}, \mnras,
  420, 1825

\bibitem[{{Kaaret} {et~al.}(2017){Kaaret}, {Feng}, \&
  {Roberts}}]{2017ARA&A..55..303K}
{Kaaret}, P., {Feng}, H., \& {Roberts}, T.~P. 2017, Annual Review of Astronomy
  and Astrophysics, 55, 303

\bibitem[{{Kellerman} {et~al.}(1989){Kellerman}, {Sramek}, {Schmidt},
  {Shaffer}, \& {Green}}]{1989AJ.....98.1195K}
{Kellerman}, K.~I., {Sramek}, R., {Schmidt}, M., {Shaffer}, D.~B., \& {Green},
  R. 1989, \aj, 98, 1195

\bibitem[{{Kelley} {et~al.}(2014){Kelley}, {Tchekhovskoy}, \&
  {Narayan}}]{2014MNRAS.445.3919K}
{Kelley}, L.~Z., {Tchekhovskoy}, A., \& {Narayan}, R. 2014, \mnras, 445, 3919

\bibitem[{{Kennicutt} \& {Evans}(2012)}]{2012ARA&A..50..531K}
{Kennicutt}, R.~C., \& {Evans}, N.~J. 2012, Annual Review of Astronomy and
  Astrophysics, 50, 531

\bibitem[{{Kinney} {et~al.}(2000){Kinney}, {Schmitt}, {Clarke}, {Pringle},
  {Ulvestad}, \& {Antonucci}}]{2000ApJ...537..152K}
{Kinney}, A.~L., {Schmitt}, H.~R., {Clarke}, C.~J., {et~al.} 2000, \apj, 537,
  152

\bibitem[{{K{\"o}rding} {et~al.}(2006){K{\"o}rding}, {Jester}, \&
  {Fender}}]{2006MNRAS.372.1366K}
{K{\"o}rding}, E.~G., {Jester}, S., \& {Fender}, R. 2006, \mnras, 372, 1366

\bibitem[{{Kormendy} \& {Ho}(2013)}]{2013ARA&A..51..511K}
{Kormendy}, J., \& {Ho}, L.~C. 2013, \araa, 51, 511

\bibitem[{{Kormendy} \& {Kennicutt}(2004)}]{2004ARA&A..42..603K}
{Kormendy}, J., \& {Kennicutt}, Robert~C., J. 2004, Annual Review of Astronomy
  and Astrophysics, 42, 603

\bibitem[{{Kronberg} \& {Sramek}(1985)}]{1985Sci...227...28K}
{Kronberg}, P.~P., \& {Sramek}, R.~A. 1985, Science, 227, 28

\bibitem[{{Kronberg} \& {Sramek}(1992)}]{1992xrea.conf..247K}
{Kronberg}, P.~P., \& {Sramek}, R.~A. 1992, in X-ray Emission from Active
  Galactic Nuclei and the Cosmic X-ray Background, 247

\bibitem[{{Ku{\'z}micz} {et~al.}(2018){Ku{\'z}micz}, {Jamrozy}, {Bronarska},
  {Janda- Boczar}, \& {Saikia}}]{2018arXiv180909008K}
{Ku{\'z}micz}, A., {Jamrozy}, M., {Bronarska}, K., {Janda- Boczar}, K., \&
  {Saikia}, D.~J. 2018, ArXiv e-prints, arXiv:1809.09008

\bibitem[{{Lacey} {et~al.}(1997){Lacey}, {Duric}, \&
  {Goss}}]{1997ApJS..109..417L}
{Lacey}, C., {Duric}, N., \& {Goss}, W.~M. 1997, The Astrophysical Journal
  Supplement Series, 109, 417

\bibitem[{{Lal} {et~al.}(2004){Lal}, {Shastri}, \&
  {Gabuzda}}]{2004A&A...425...99L}
{Lal}, D.~V., {Shastri}, P., \& {Gabuzda}, D.~C. 2004, \aap, 425, 99

\bibitem[{{Lanzuisi} {et~al.}(2016){Lanzuisi}, {Perna}, {Comastri}, {Cappi},
  {Dadina}, {Marinucci}, {Masini}, {Matt}, {Vagnetti}, {Vignali}, {Ballantyne},
  {Bauer}, {Boggs}, {Brandt}, {Brusa}, {Christensen}, {Craig}, {Fabian},
  {Farrah}, {Hailey}, {Harrison}, {Luo}, {Piconcelli}, {Puccetti}, {Ricci},
  {Saez}, {Stern}, {Walton}, \& {Zhang}}]{2016A&A...590A..77L}
{Lanzuisi}, G., {Perna}, M., {Comastri}, A., {et~al.} 2016, \aap, 590, A77

\bibitem[{{Laor} {et~al.}(2019){Laor}, {Baldi}, \&
  {Behar}}]{2019MNRAS.482.5513L}
{Laor}, A., {Baldi}, R.~D., \& {Behar}, E. 2019, \mnras, 482, 5513

\bibitem[{{Laor} \& {Behar}(2008)}]{2008MNRAS.390..847L}
{Laor}, A., \& {Behar}, E. 2008, \mnras, 390, 847

\bibitem[{{Levan} {et~al.}(2011){Levan}, {Tanvir}, {Cenko}, {Perley},
  {Wiersema}, {Bloom}, {Fruchter}, {Postigo}, {O'Brien}, {Butler}, {van der
  Horst}, {Leloudas}, {Morgan}, {Misra}, {Bower}, {Farihi}, {Tunnicliffe},
  {Modjaz}, {Silverman}, {Hjorth}, {Th{\"o}ne}, {Cucchiara}, {Cer{\'o}n},
  {Castro-Tirado}, {Arnold}, {Bremer}, {Brodie}, {Carroll}, {Cooper}, {Curran},
  {Cutri}, {Ehle}, {Forbes}, {Fynbo}, {Gorosabel}, {Graham}, {Hoffman},
  {Guziy}, {Jakobsson}, {Kamble}, {Kerr}, {Kasliwal}, {Kouveliotou},
  {Kocevski}, {Law}, {Nugent}, {Ofek}, {Poznanski}, {Quimby}, {Rol},
  {Romanowsky}, {S{\'a}nchez-Ram{\'\i}rez}, {Schulze}, {Singh}, {van
  Spaandonk}, {Starling}, {Strom}, {Tello}, {Vaduvescu}, {Wheatley}, {Wijers},
  {Winters}, \& {Xu}}]{2011Sci...333..199L}
{Levan}, A.~J., {Tanvir}, N.~R., {Cenko}, S.~B., {et~al.} 2011, Science, 333,
  199

\bibitem[{{Lietzen} {et~al.}(2011){Lietzen}, {Hein{\"a}m{\"a}ki}, {Nurmi},
  {Liivam{\"a}gi}, {Saar}, {Tago}, {Takalo}, \&
  {Einasto}}]{2011A&A...535A..21L}
{Lietzen}, H., {Hein{\"a}m{\"a}ki}, P., {Nurmi}, P., {et~al.} 2011, \aap, 535,
  A21

\bibitem[{{Lisenfeld} \& {V{\"o}lk}(2000)}]{2000A&A...354..423L}
{Lisenfeld}, U., \& {V{\"o}lk}, H.~J. 2000, \aap, 354, 423

\bibitem[{{Lu} \& {Yu}(1999)}]{1999ApJ...526L...5L}
{Lu}, Y., \& {Yu}, Q. 1999, \apjl, 526, L5

\bibitem[{{Marconi} {et~al.}(2004){Marconi}, {Risaliti}, {Gilli}, {Hunt},
  {Maiolino}, \& {Salvati}}]{2004MNRAS.351..169M}
{Marconi}, A., {Risaliti}, G., {Gilli}, R., {et~al.} 2004, \mnras, 351, 169

\bibitem[{{Mart{\'\i}nez-Paredes} {et~al.}(2017){Mart{\'\i}nez-Paredes},
  {Aretxaga}, {Alonso-Herrero}, {Gonz{\'a}lez-Mart{\'\i}n},
  {Lop{\'e}z-Rodr{\'\i}guez}, {Ramos Almeida}, {Asensio Ramos}, {Diaz Santos},
  {Elitzur}, {Esquej}, {Hern{\'a}n-Caballero}, {Ichikawa}, {Nikutta},
  {Packham}, {Pereira-Santaella}, \& {Telesco}}]{2017MNRAS.468....2M}
{Mart{\'\i}nez-Paredes}, M., {Aretxaga}, I., {Alonso-Herrero}, A., {et~al.}
  2017, \mnras, 468, 2

\bibitem[{{McKinney} {et~al.}(2015){McKinney}, {Dai}, \&
  {Avara}}]{2015MNRAS.454L...6M}
{McKinney}, J.~C., {Dai}, L., \& {Avara}, M.~J. 2015, \mnras, 454, L6

\bibitem[{{McLure} \& {Dunlop}(2004)}]{2004MNRAS.352.1390M}
{McLure}, R.~J., \& {Dunlop}, J.~S. 2004, \mnras, 352, 1390

\bibitem[{{McMullin} {et~al.}(2007){McMullin}, {Waters}, {Schiebel}, {Young},
  \& {Golap}}]{2007ASPC..376..127M}
{McMullin}, J.~P., {Waters}, B., {Schiebel}, D., {Young}, W., \& {Golap}, K.
  2007, in Astronomical Data Analysis Software and Systems XVI, Vol. 376, 127

\bibitem[{{Meier}(1996)}]{1996ApJ...459..185M}
{Meier}, D. 1996, \apj, 459, 185

\bibitem[{{Merloni} {et~al.}(2003){Merloni}, {Heinz}, \& {di
  Matteo}}]{2003MNRAS.345.1057M}
{Merloni}, A., {Heinz}, S., \& {di Matteo}, T. 2003, \mnras, 345, 1057

\bibitem[{{Middleton} {et~al.}(2018){Middleton}, {Walton}, {Alston}, {Dauser},
  {Eikenberry}, {Jiang}, {Fabian}, {Fuerst}, {Brightman}, {Marshall}, {Parker},
  {Pinto}, {Harrison}, {Bachetti}, {Altamirano}, {Bird}, {Perez},
  {Miller-Jones}, {Charles}, {Boggs}, {Christensen}, {Craig}, {Forster},
  {Grefenstette}, {Hailey}, {Madsen}, {Stern}, \&
  {Zhang}}]{2018arXiv181010518M}
{Middleton}, M.~J., {Walton}, D.~J., {Alston}, W., {et~al.} 2018, arXiv
  e-prints, arXiv:1810.10518

\bibitem[{{Miller} \& {Scalo}(1979)}]{1979ApJS...41..513M}
{Miller}, G.~E., \& {Scalo}, J.~M. 1979, The Astrophysical Journal Supplement
  Series, 41, 513

\bibitem[{{Mirabel} \& {Rodr{\'\i}guez}(1994)}]{1994Natur.371...46M}
{Mirabel}, I.~F., \& {Rodr{\'\i}guez}, L.~F. 1994, \nat, 371, 46

\bibitem[{{Mu{\~n}oz Mar{\'\i}n} {et~al.}(2007){Mu{\~n}oz Mar{\'\i}n},
  {Gonz{\'a}lez Delgado}, {Schmitt}, {Cid Fernandes}, {P{\'e}rez},
  {Storchi-Bergmann}, {Heckman}, \& {Leitherer}}]{2007AJ....134..648M}
{Mu{\~n}oz Mar{\'\i}n}, V.~M., {Gonz{\'a}lez Delgado}, R.~M., {Schmitt}, H.~R.,
  {et~al.} 2007, \aj, 134, 648

\bibitem[{{Muxlow} {et~al.}(1994){Muxlow}, {Pedlar}, {Wilkinson}, {Axon},
  {Sanders}, \& {de Bruyn}}]{1994MNRAS.266..455M}
{Muxlow}, T.~W.~B., {Pedlar}, A., {Wilkinson}, P.~N., {et~al.} 1994, \mnras,
  266, 455

\bibitem[{{Narayan} {et~al.}(2003){Narayan}, {Igumenshchev}, \&
  {Abramowicz}}]{2003PASJ...55L..69N}
{Narayan}, R., {Igumenshchev}, I.~V., \& {Abramowicz}, M.~A. 2003, Publications
  of the Astronomical Society of Japan, 55, L69

\bibitem[{{Neilsen} \& {Lee}(2009)}]{2009Natur.458..481N}
{Neilsen}, J., \& {Lee}, J.~C. 2009, \nat, 458, 481

\bibitem[{{Ohsuga} \& {Mineshige}(2011)}]{2011ApJ...736....2O}
{Ohsuga}, K., \& {Mineshige}, S. 2011, \apj, 736, 2

\bibitem[{{Okuda} {et~al.}(2005){Okuda}, {Teresi}, {Toscano}, \&
  {Molteni}}]{2005MNRAS.357..295O}
{Okuda}, T., {Teresi}, V., {Toscano}, E., \& {Molteni}, D. 2005, \mnras, 357,
  295

\bibitem[{{Osterbrock} \& {Pogge}(1985)}]{1985ApJ...297..166O}
{Osterbrock}, D.~E., \& {Pogge}, R.~W. 1985, \apj, 297, 166

\bibitem[{{Padovani} {et~al.}(2011){Padovani}, {Miller}, {Kellermann},
  {Mainieri}, {Rosati}, \& {Tozzi}}]{2011ApJ...740...20P}
{Padovani}, P., {Miller}, N., {Kellermann}, K.~I., {et~al.} 2011, \apj, 740, 20

\bibitem[{{Panessa} {et~al.}(2019){Panessa}, {Baldi}, {Laor}, {Padovani},
  {Behar}, \& {McHardy}}]{2019NatAs...3..387P}
{Panessa}, F., {Baldi}, R.~D., {Laor}, A., {et~al.} 2019, Nature Astronomy, 3,
  387

\bibitem[{{P{\'e}rez-Torres} {et~al.}(2009){P{\'e}rez-Torres},
  {Romero-Ca{\~n}izales}, {Alberdi}, \& {Polatidis}}]{2009A&A...507L..17P}
{P{\'e}rez-Torres}, M.~A., {Romero-Ca{\~n}izales}, C., {Alberdi}, A., \&
  {Polatidis}, A. 2009, \aap, 507, L17

\bibitem[{{Perley} \& {Butler}(2013)}]{2013ApJS..204...19P}
{Perley}, R.~A., \& {Butler}, B.~J. 2013, The Astrophysical Journal Supplement
  Series, 204, 19

\bibitem[{{Rabidoux} {et~al.}(2014){Rabidoux}, {Pisano}, {Kepley}, {Johnson},
  \& {Balser}}]{2014ApJ...780...19R}
{Rabidoux}, K., {Pisano}, D.~J., {Kepley}, A.~A., {Johnson}, K.~E., \&
  {Balser}, D.~S. 2014, \apj, 780, 19

\bibitem[{{Raginski} \& {Laor}(2016)}]{2016MNRAS.459.2082R}
{Raginski}, I., \& {Laor}, A. 2016, \mnras, 459, 2082

\bibitem[{{Reines} {et~al.}(2020){Reines}, {Condon}, {Darling}, \&
  {Greene}}]{2020ApJ...888...36R}
{Reines}, A.~E., {Condon}, J.~J., {Darling}, J., \& {Greene}, J.~E. 2020, \apj,
  888, 36.
\newblock \doarXiv{1909.04670}

\bibitem[{{Ricci} {et~al.}(2013){Ricci}, {Paltani}, {Ueda}, \&
  {Awaki}}]{2013MNRAS.435.1840R}
{Ricci}, C., {Paltani}, S., {Ueda}, Y., \& {Awaki}, H. 2013, \mnras, 435, 1840

\bibitem[{{Richards} \& {Lister}(2015)}]{2015ApJ...800L...8R}
{Richards}, J.~L., \& {Lister}, M.~L. 2015, \apj, 800, L8

\bibitem[{{Roberts} {et~al.}(2016){Roberts}, {Middleton}, {Sutton}, {Mezcua},
  {Walton}, \& {Heil}}]{2016AN....337..534R}
{Roberts}, T.~P., {Middleton}, M.~J., {Sutton}, A.~D., {et~al.} 2016,
  Astronomische Nachrichten, 337, 534

\bibitem[{{Runnoe} {et~al.}(2012){Runnoe}, {Brotherton}, \&
  {Shang}}]{2012MNRAS.426.2677R}
{Runnoe}, J.~C., {Brotherton}, M.~S., \& {Shang}, Z. 2012, \mnras, 426, 2677

\bibitem[{{Schawinski} {et~al.}(2015){Schawinski}, {Koss}, {Berney}, \&
  {Sartori}}]{2015MNRAS.451.2517S}
{Schawinski}, K., {Koss}, M., {Berney}, S., \& {Sartori}, L.~F. 2015, \mnras,
  451, 2517

\bibitem[{{Schmitt} {et~al.}(2001){Schmitt}, {Ulvestad}, {Antonucci}, \&
  {Kinney}}]{2001ApJS..132..199S}
{Schmitt}, H.~R., {Ulvestad}, J.~S., {Antonucci}, R.~R.~J., \& {Kinney}, A.~L.
  2001, The Astrophysical Journal Supplement Series, 132, 199

\bibitem[{{Seaquist} \& {Odegard}(1991)}]{1991ApJ...369..320S}
{Seaquist}, E.~R., \& {Odegard}, N. 1991, \apj, 369, 320

\bibitem[{{Shepherd} {et~al.}(1994){Shepherd}, {Pearson}, \&
  {Taylor}}]{1994BAAS...26..987S}
{Shepherd}, M.~C., {Pearson}, T.~J., \& {Taylor}, G.~B. 1994, in Bulletin of
  the American Astronomical Society, Vol.~26, 987--989

\bibitem[{{Sikora} {et~al.}(2007){Sikora}, {Stawarz}, \&
  {Lasota}}]{2007ApJ...658..815S}
{Sikora}, M., {Stawarz}, {\L}., \& {Lasota}, J.-P. 2007, \apj, 658, 815

\bibitem[{{S{\k{a}}dowski} \& {Narayan}(2015)}]{2015MNRAS.453.3213S}
{S{\k{a}}dowski}, A., \& {Narayan}, R. 2015, \mnras, 453, 3213

\bibitem[{{Sutton} {et~al.}(2013){Sutton}, {Roberts}, \&
  {Middleton}}]{2013MNRAS.435.1758S}
{Sutton}, A.~D., {Roberts}, T.~P., \& {Middleton}, M.~J. 2013, \mnras, 435,
  1758

\bibitem[{{Takeo} {et~al.}(2018){Takeo}, {Inayoshi}, {Ohsuga}, {Takahashi}, \&
  {Mineshige}}]{2018MNRAS.476..673T}
{Takeo}, E., {Inayoshi}, K., {Ohsuga}, K., {Takahashi}, H.~R., \& {Mineshige},
  S. 2018, \mnras, 476, 673

\bibitem[{{Takeuchi} {et~al.}(2009){Takeuchi}, {Mineshige}, \&
  {Ohsuga}}]{2009PASJ...61..783T}
{Takeuchi}, S., {Mineshige}, S., \& {Ohsuga}, K. 2009, Publications of the
  Astronomical Society of Japan, 61, 783

\bibitem[{{Tarchi} {et~al.}(2000){Tarchi}, {Neininger}, {Greve}, {Klein},
  {Garrington}, {Muxlow}, {Pedlar}, \& {Glendenning}}]{2000A&A...358...95T}
{Tarchi}, A., {Neininger}, N., {Greve}, A., {et~al.} 2000, \aap, 358, 95

\bibitem[{{Ulvestad} \& {Antonucci}(1997)}]{1997ApJ...488..621U}
{Ulvestad}, J.~S., \& {Antonucci}, R. R.~J. 1997, \apj, 488, 621

\bibitem[{{Ulvestad} {et~al.}(2005){Ulvestad}, {Antonucci}, \&
  {Barvainis}}]{2005ApJ...621..123U}
{Ulvestad}, J.~S., {Antonucci}, R. R.~J., \& {Barvainis}, R. 2005, \apj, 621,
  123

\bibitem[{{Ulvestad} {et~al.}(1995){Ulvestad}, {Antonucci}, \&
  {Goodrich}}]{1995AJ....109...81U}
{Ulvestad}, J.~S., {Antonucci}, R. R.~J., \& {Goodrich}, R.~W. 1995, \aj, 109,
  81

\bibitem[{{Varenius} {et~al.}(2014){Varenius}, {Conway}, {Mart{\'\i}-Vidal},
  {Aalto}, {Beswick}, {Costagliola}, \& {Kl{\"o}ckner}}]{2014A&A...566A..15V}
{Varenius}, E., {Conway}, J.~E., {Mart{\'\i}-Vidal}, I., {et~al.} 2014, \aap,
  566, A15

\bibitem[{{Vasudevan} \& {Fabian}(2007)}]{2007MNRAS.381.1235V}
{Vasudevan}, R.~V., \& {Fabian}, A.~C. 2007, \mnras, 381, 1235

\bibitem[{{Volonteri} \& {Rees}(2005)}]{2005ApJ...633..624V}
{Volonteri}, M., \& {Rees}, M.~J. 2005, \apj, 633, 624

\bibitem[{{Volonteri} {et~al.}(2015){Volonteri}, {Silk}, \&
  {Dubus}}]{2015ApJ...804..148V}
{Volonteri}, M., {Silk}, J., \& {Dubus}, G. 2015, \apj, 804, 148

\bibitem[{{Wang} {et~al.}(2013){Wang}, {Du}, {Valls-Gabaud}, {Hu}, \&
  {Netzer}}]{2013PhRvL.110h1301W}
{Wang}, J.-M., {Du}, P., {Valls-Gabaud}, D., {Hu}, C., \& {Netzer}, H. 2013,
  Physical Review Letters, 110, 081301

\bibitem[{{Wu} {et~al.}(2018){Wu}, {Coughlin}, \&
  {Nixon}}]{2018MNRAS.478.3016W}
{Wu}, S., {Coughlin}, E.~R., \& {Nixon}, C. 2018, \mnras, 478, 3016

\bibitem[{{Yang} {et~al.}(2018){Yang}, {An}, {Zheng}, {Baan}, {Paragi},
  {Mohan}, {Zhang}, \& {Liu}}]{2018arXiv181006067Y}
{Yang}, J., {An}, T., {Zheng}, F., {et~al.} 2018, ArXiv e-prints,
  arXiv:1810.06067

\bibitem[{{Zauderer} {et~al.}(2011){Zauderer}, {Berger}, {Soderberg}, {Loeb},
  {Narayan}, {Frail}, {Petitpas}, {Brunthaler}, {Chornock}, {Carpenter},
  {Pooley}, {Mooley}, {Kulkarni}, {Margutti}, {Fox}, {Nakar}, {Patel},
  {Volgenau}, {Culverhouse}, {Bietenholz}, {Rupen}, {Max-Moerbeck}, {Readhead},
  {Richards}, {Shepherd}, {Storm}, \& {Hull}}]{2011Natur.476..425Z}
{Zauderer}, B.~A., {Berger}, E., {Soderberg}, A.~M., {et~al.} 2011, \nat, 476,
  425

\bibitem[{{Zhou} {et~al.}(2007){Zhou}, {Wang}, {Yuan}, {Shan}, {Komossa}, {Lu},
  {Liu}, {Xu}, {Bai}, \& {Jiang}}]{2007zhou}
{Zhou}, H., {Wang}, T., {Yuan}, W., {et~al.} 2007, ApJL, 658, L13

\end{thebibliography}



\end{document}